\documentclass[fleqn,10pt,twocolumn,]{wlscirep}
\usepackage[utf8]{inputenc}
\usepackage[T1]{fontenc}
\usepackage{mathtools,multirow,amssymb,commath,siunitx,gensymb}
\usepackage{placeins} 

\usepackage[final]{pdfpages} 

\title{Phase stability and mechanical property trends for MAB phases by high-throughput ab initio calculations}

\author[1,2,*]{Nikola Koutn\'a}
\author[2]{Lars Hultman}
\author[1]{Paul H. Mayrhofer}
\author[2]{Davide G. Sangiovanni}

\affil[1]{Institute of Materials Science and Technology, TU Wien, Getreidemarkt 9, A-1060 Vienna, Austria}

\affil[2]{Department of Physics, Chemistry, and Biology (IFM), Link\"{o}ping University, SE-58183 Link\"{o}ping, Sweden}

\affil[*]{nikola.koutna@tuwien.ac.at; nikola.koutna@liu.se}

\begin{abstract}
MAB phases (MABs) are atomically-thin laminates of ceramic/metallic-like layers, having made a breakthrough in the development of 2D materials. 
Though theoretically offering a vast chemical and phase space, relatively few MABs have yet been synthesised. 
To guide experiments, we perform a systematic high-throughput {\it{ab initio}} screening of MABs that combine group 4--7 transition metals (M); Al, Si, Ga, Ge, or In (A); and boron (B) focusing on their phase stability trends and mechanical properties.
Considering the 1:1:1, 2:1:1, 2:1:2, 3:1:2, 3:1:3, and 3:1:4 M:A:B ratios and 10 phase prototypes, possible stabilisation of a single-phase compound for each elemental combination is assessed through formation energy spectra of the competing mechanically and dynamically stable MABs.
Based on the volumetric proximity of energetically-close phases, we identify systems in which volume-changing deformations may facilitate transformation toughening.
Subsequently, chemistry- and phase-structure-related trends in the elastic stiffness and ductility are predicted using elastic-constants-based descriptors.
The analysis of directional Cauchy pressures and Young's moduli allows comparing mechanical response parallel and normal to M--B/A layers.  
Among the suggested most promising MABs are Nb$_3$AlB$_4$, Cr$_2$SiB$_2$, Mn$_2$SiB$_2$ or the already synthesised MoAlB.
\end{abstract}

\begin{document}

\flushbottom
\maketitle

\thispagestyle{empty}


\section{Introduction}
MAB phases (MABs) are atomically-laminated borides in which hard ceramic-like transition metal M--B layers alternate with relatively softer metallic-like mono- or bilayers of an A-element (typically Al, Si, Ga or In)\cite{kota2020progress,wang2019discovery}.
Though discovered already in the 60s\cite{aronsson1960x}, MABs have recently made a  breakthrough in the development of 2D materials for new-generation nanodevices\cite{alameda2018topochemical,zhou2021boridene,yang2020mbenes}.
Offering an interesting combination of mechanical, magnetocaloric, and catalytic properties, high-temperature oxidation resistance, as well as damage and radiation tolerance, MABs are prominent candidates for applications in the fields of protective and wear-resistant coatings, magnetic cooling, electrocatalysis or electrochemical sensing, or radiation shielding\cite{siriwardane2020revealing,kota2020progress,rosenkranz2022mab,zhang2023experimental}.

With typical formula M$_{n+1}$AB$_{2n}$ ($n=\{1,2,3\}$)\cite{carlsson2022theoretical} and possible structures with hexagonal\cite{wang2019discovery} or orthorhombic\cite{kota2020progress} symmetry, MABs theoretically provide a vast chemical and phase space.
However, relatively few material systems have been achieved experimentally: mostly bulk polycrystals (Ti$_2$InB$_2$\cite{wang2019discovery}, Cr$_2$AlB$_2$\cite{berastegui2020magnetron,kota2018magnetic,ade2015ternary}, Cr$_3$AlB$_4$\cite{kota2018magnetic,ade2015ternary}, Cr$_4$AlB$_6$\cite{ade2015ternary}, MoAlB\cite{zhang2023experimental,chen2019compressive}, WAlB\cite{zhang2023experimental,roy2023low}, Fe$_2$AlB$_2$\cite{liu2018rapid}, Mn$_2$AlB$_2$\cite{kota2018synthesis}), and only one thin film  (MoAlB\cite{achenbach2019synthesis,evertz2021low,sahu2022defects}).
Thus, further development of both bulk and thin film MAB phases calls for a systematic computational screening across a relevant subspace of the periodic table, in particular, predicting trends in the phase stability and structure--property relationships for various M and A combinations.

In terms of phase stability predictions, solid work has already been done using the {\it{ab initio}} density functional theory (DFT) framework (see, e.g.,~Refs.~\cite{dahlqvist2020theoretical,khazaei2019novel,carlsson2022theoretical,siriwardane2020revealing,shen2021designing,dahlqvist2021predictions,dahlqvist2022chemical}), or machine-learning based approaches\cite{sun2023accelerating,li2023high}.
In particular, Khazaei et al.\cite{khazaei2019novel} studied stability of the orthorhombic M$_2$AlB$_2$, MAlB, M$_3$AlB$_4$, and M$_4$AlB$_6$ systems with M from the group 3--6 transition metals, considering the decomposition into competing M--B, M--Al, and M--Al--B compounds.  
Siriwardane et al.\cite{siriwardane2020revealing} suggested that stability of a MAB phase for a given M decreases for A changing from Al$\to$Ga$\to$In$\to$Tl, i.e., with increasing atomic number of the A element. 
This observation was correlated with increased M--A and B--A bond lengths, causing a decreased ionicity.  
Later, Carlsson et al.\cite{carlsson2022theoretical} screened orthorhombic and hexagonal MAB, M$_2$AB$_2$, M$_3$AB$_4$, M$_4$AB$_4$, and M$_4$AB$_6$ phases with M from the group 3--6 transition metals or Mn, Fe, Co, and A$=\{$Al, Ga, In\}, confirming thermodynamic stability of 7 previously synthesised MABs, predicting 3 additional ones to be stable and 23 nearly stable.
Furthermore, the authors hypothesised on preferential orthorhombic/hexagonal symmetry for MAB phases containing Al and \{Ga, In\}, respectively.

Most {\it{ab initio}} investigations have been directed to identification of new stable MABs, while only few aimed on systematic predictions of materials' properties and their relationships to phase prototypes and/or elemental composition\cite{li2023high,lind2021plane,liu2020new,qi2023stability}. 
For purposes of this work, we focus on studies addressing mechanical behaviour\cite{liu2020new,zhou2017electrical,dai2018first,lind2021plane,dai2017easily}.
Within the DFT framework, this is typically realised by calculating the elastic constants ($C_{ij}$) and deriving phenomenological strength and ductility indicators: the Young's modulus, shear-to-bulk modulus ratio, or Cauchy pressure (widely accepted trend-givers in the family of ceramics\cite{moraes2018ab,koutna2021high,balasubramanian2018valence,sangiovanni2010electronic,gu2021sorting}). 
Employing $C_{ij}$-based indicators, Liu et al.\cite{liu2020new} proposed TcAlB, NbAlB, WAlB, Tc$_2$AlB$_2$, Co$_2$AlB2, and Ni$_2$AlB$_2$ to be the most ductile and Mo$_2$AlB$_2$ with W$_2$AlB$_2$ to be the stiffest MABs of the MAB and M$_2$AB$_2$ phase prototypes, considering M from the group 3--5 transition metals and A$=$Al.
Going beyond calculations of elastic constants, Dai et al.\cite{dai2017easily} modelled (0~K) shear deformation of (CrB$_2$)$_n$CrAl, $n=\{1,2,3\}$, suggesting that tiltable B--A--B bonds can release shear strain in weaker A layers, hence, contribute to high fracture toughness and damage tolerance.

Experimental investigations on MABs' mechanical response have most often concerned Young's modulus measurements (e.g., MoAlB\cite{achenbach2019synthesis,evertz2021low} and Mn$_2$AlB$_2$\cite{kota2018synthesis}) whereas toughness-related quantities have been significantly less researched.
For example, Fe$_2$AlB$_2$\cite{li2017rapid} exhibited $K_{1C}=5.4\pm0.2$~MPa$\sqrt{\text{m}}$, which exceeds typical values for transition metal diborides\cite{fuger2023tissue,fuger2019influence,hahn2023unraveling}).
Recently, crack deflection and crack healing behaviour have been reported for MoAlB\cite{lu2019crack,lu2019thermal} and Fe$_2$AlB$_2$\cite{bai2019high}, further motivating the need for theory-based understanding of MABs' mechanical response in relation to their chemistry and phase structure.

Our study uses high-throughput DFT calculations to map phase stability trends, structural and mechanical properties of MAB phases containing group 4--7 transition metals, $\text{M}=($Ti, Zr, Hf; V, Nb, Ta; Cr, Mo, W; Mn, Tc, Re), $\text{A}=($Al, Ga, In, Si, Ge), and boron (B). 
For each elemental combination, 10 phase prototypes with various M:A:B ratios are considered.
These include experimentally known MAB and MAX (X$=$C, N) phases\cite{kota2020progress,sokol2019chemical,wang2019discovery}, or are inspired by common transition metal (di)boride structures, exhibiting intrinsically layered character but not yet regarded as possible MAB structures.
Our initial screening concerns formation energy, mechanical, and dynamical stability calculations, identification of the most favourable structure, and prediction of a single-phase MABs synthesisability.
For systems passing our stability criteria, chemistry- and phase-structure-related trends in elastic stiffness and ductility are predicted, including both the polycrystalline approximates and directional values parallel and normal to the metal/ceramic layers.
Finally, the most promising candidates for the synthesis of novel MABs are suggested.

\section{Methods}
Density Functional Theory (DFT) calculations were performed using the Vienna Ab-initio Simulation Package (VASP)\cite{Kresse1996Efficient, Kresse1999From} together with the projector augmented plane-wave (PAW) method\cite{Kohn1965Self} and the Perdew-Burke-Ernzerhof (PBE) generalised gradient approximation\cite{PhysRevLett.77.3865}.
Following convergence tests, the plane-wave cutoff energy was set to 600~eV and the $\Gamma$--centred $k$-point mesh of the Brillouin-zone was automatically generated with a length parameter ($R_k$) of 60~\AA\ (i.e., $k$-points separated by $1/60$~\AA$^{-1}$ along each $b$ vector). 

\begin{figure}[h!t!]
    \centering
    \includegraphics[width=1\columnwidth]{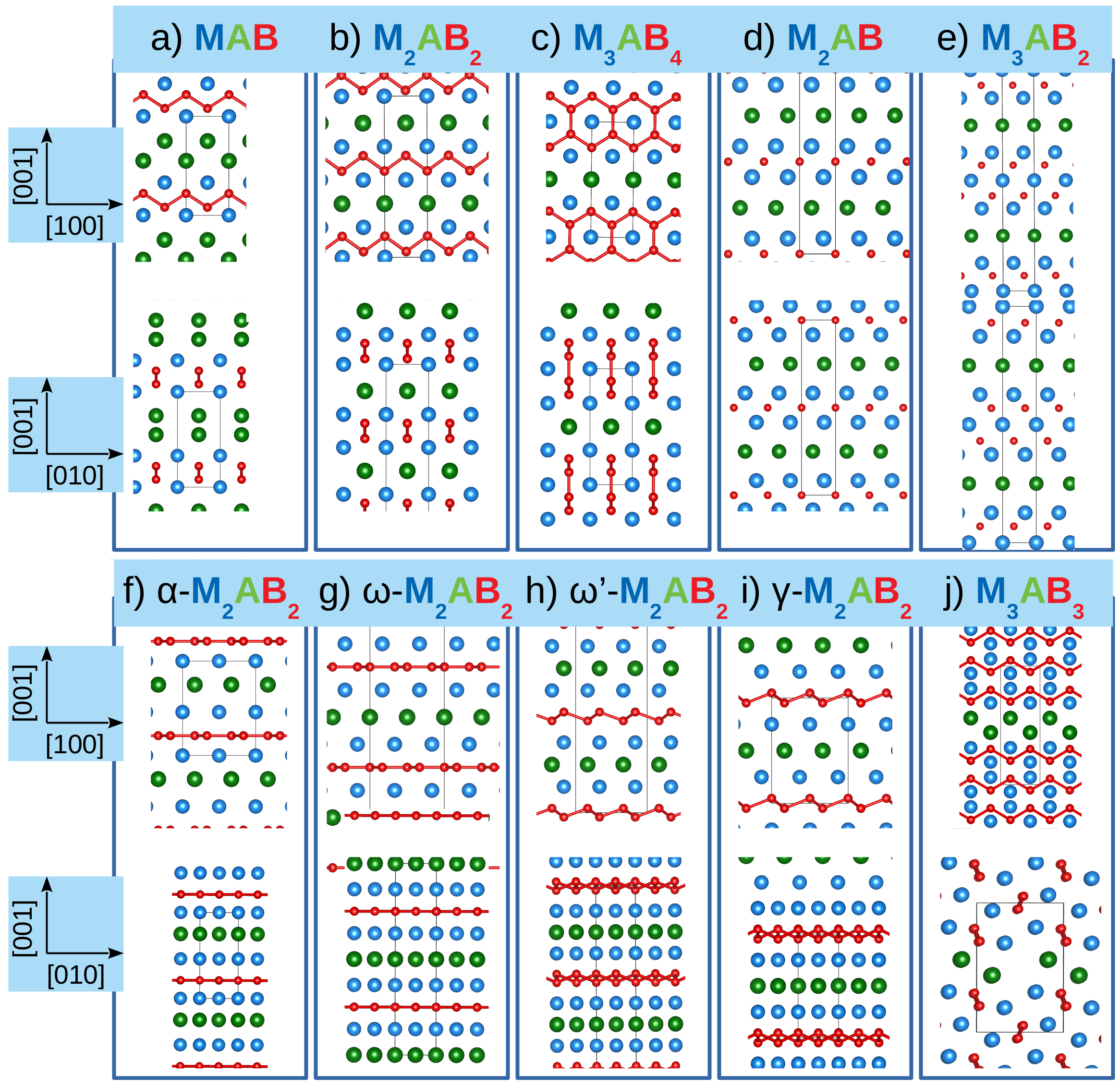}
    \caption{\small 
    {\bf{Snapshots of the here considered MAB phase prototypes}}. 
    (a--c) The {\bf{MAB}}, {\bf{M$_2$AB$_2$}}, {\bf{M$_3$AB$_4$}} are known MAB phase prototypes~\cite{kota2020progress} experimentally achieved for, e.g., MoAlB\cite{achenbach2019synthesis}, Mn$_2$AlB$_2$\cite{kota2018synthesis}, and Cr$_3$AlB$_4$\cite{kota2018magnetic,ade2015ternary}.
    (d--e) The {\bf{M$_2$AB}} and {\bf{M$_3$AB$_2$}} are experimentally known MAX ($X=$ C, N) phase prototypes\cite{sokol2019chemical}, where the M$_2$AB structure has been considered for MABs in a recent DFT study\cite{aydin2020211}.
    (f--j) The {\bf{$\alpha$-M$_2$AB$_2$}}, {\bf{$\omega$-M$_2$AB$_2$}}, {\bf{$\omega$'-M$_2$AB$_2$}}, {\bf{$\gamma$-M$_2$AB$_2$}}, and {\bf{M$_3$AB$_3$}} are hand-designed based on common structures of transition metal borides (TMBs).
    The $\alpha$-M$_2$AB$_2$ is based on the hexagonal $\alpha$-AlB$_2$ type phase ($P6/mmm$; typical for the group 4--6 TMB$_2$s\cite{magnuson2021review}) and has been reported for a bulk Ti$_2$InB$_2$ MAB phase\cite{wang2019discovery}. 
    The $\omega$-M$_2$AB$_2$, $\omega$'-M$_2$AB$_2$, and $\gamma$-M$_2$AB$_2$ are based on the $\omega$-WB$_2$ type phase (P6$_3$/mmc; ABBA stacking) and the $\gamma$-ReB$_2$ type phase (P6$_3$/mmc; BABA stacking), see Ref.~\cite{leiner2023energetics}. 
    The difference between $\omega$-M$_2$AB$_2$ and $\omega$'-M$_2$AB$_2$ is that in the former (latter), Al replaces the flat (puckered) B sheets while the puckered (flat) sheets remain B and form the ceramic M--B layer. 
    The M$_3$AB$_3$ prototype is based on the structure of TiB (Pnma)\cite{panda2006first,jain2013commentary}. 
    Other structures of common boride-based ceramics (e.g., Ti$_3$B$_4$ and Ti$_2$B) inspired hypothetical phase prototypes of additional MABs, however, were found irrelevant due to their high formation energies, thus excluded.
    }
\label{FIG: prototypes}
\end{figure}

In total 60 atomically-laminated M--A--B systems were modelled, including all combinations of $\text{M}=($Ti, Zr, Hf; V, Nb, Ta; Cr, Mo, W; Mn, Tc, Re), $\text{A}=($Al, Ga, In, Si, Ge) and boron (B).
For each elemental combination, we considered the 1:1:1, 2:1:1, 2:1:2, 3:1:2, 3:1:3, and 3:1:4 ratio between M, A, and boron, and 10 MAB phase prototypes given below and schematically depicted in Fig.~\ref{FIG: prototypes} (the thereby established notation will be used throughout this work):
\begin{itemize}\setlength\itemsep{0em}
    \item For the {\bf{1:1:1 chemistry}}: {\bf{MAB}} (orthorhombic with a space group (s.g.) $Cmcm$; a 12-atom simulation cell).  

    \item For the {\bf{2:1:1 chemistry}}: {\bf{M$_2$AB}} (hexagonal with a s.g. $P6_3/mmc$; an 8-atom simulation cell).
    
    \item For the {\bf{2:1:2 chemistry}}: {\bf{M$_2$AB$_2$}} (orthorhombic with a s.g. $Cmmm$; a 10-atom simulation cell), $\alpha$-M$_2$AB$_2$ (hexagonal with a s.g. $P\overline{6}m2$; a 10-atom simulation cell), {\bf{$\omega$-M$_2$AB$_2$}} (hexagonal with a s.g. P6$_3$/mmc, a 20-atom simulation cell), and {\bf{$\omega$'-M$_2$AB$_2$}} (hexagonal with a s.g. P6$_3$/mmc, a 20-atom simulation cell), {\bf{$\gamma$-M$_2$AB$_2$}} (hexagonal with a s.g. P6$_3$/mmc, a 10-atom simulation cell).

    \item For the {\bf{3:1:2 chemistry}}: {\bf{M$_3$AB$_2$}} (hexagonal with a s.g. $P6_3/mmc$; a 12-atom simulation cell).
    
    \item For the {\bf{3:1:3 chemistry}}: {\bf{M$_3$AB$_3$}} (orthorhombic with a s.g. Pnma; a 14-atom simulation cell).
    
    \item For the {\bf{3:1:4 chemistry}}: {\bf{M$_3$AB$_4$}} (orthorhombic with a s.g. Pmmm; an 8-atom simulation cell).   
\end{itemize}

The above-described phase prototypes were fully relaxed for all combinations of M and A elements (in total 600 MABs), until forces on ions did not exceed 0.005\;eV/\AA\ and total energies were converged up to 10$^{-5}$~eV/supercell.
Cr- and Mn-based MAB phases have been considered in their non-magnetic state.
Subsequently, relative chemical stability was estimated by the energy of formation  
\begin{equation}
E_f=\frac{1}{\sum_s n_s}\bigg(E_{\text{tot}}-\sum_s n_s\mu_s\bigg)\ ,
\label{Eq: Ef}
\end{equation} 
where $E_{\mathrm{tot}}$ is the total energy of the simulation cell (from the last ionic step of a structure relaxation), $n_s$ and $\mu_s$ are the number of atoms and the chemical potential, respectively, of a species $s$. 
Chemical potentials, $\mu_s$, were conventionally set to the total energy per atom of the ground-state structure for the respective element, from Material's project \cite{jain2013commentary} that is fcc-Al; bcc-V, -Nb, -Ta, -Cr, -Mo, -W, -Mn; diamond-Si, -Ge; tetragonal-In; orthorhombic-Ga; rhomboedral-B; and hcp-Ti, -Zr, -Hf, -Tc, -Re. 

The stress-strain method\cite{le2002symmetry, le2001symmetry, Yu2010-vr} was used to calculate fourth order elasticity tensors (according to Hooke's law), which were projected onto symmetric $6\times6$ matrices of elastic constants, $C_{ij}$, using the Voigt notation.
Positive definiteness of the $C_{ij}$ matrix was verified in order to determine mechanical stability of the corresponding structure\cite{mouhat2014necessary}. 
For mechanically stable MABs, their dynamical stability was assessed based on the corresponding phonon spectra: by checking for no imaginary phonon modes (i.e. non-zero phonon density of states only in the positive frequency region). 
The phonon spectra were obtained with the aid of the Phonopy package\cite{togo2008first}, using the finite displacement method with the default displacement of 0.01~\AA\ and 2$\times$2$\times$2 replicas of the fully relaxed MAB structures (supercells with 64--160 atoms). 
The supercells size effects have been tested for several cases, indicating that the chosen supercells are sufficient to verify dynamical stability. 

The $C_{ij}$ matrices of MABs fulfilling conditions for mechanical and dynamical stability were further post-processed to estimate mechanical properties.
Imposing the macroscopic symmetry, the matrices were projected on those of a hexagonal or an orthorhombic system, yielding 5 and 9 independent elastic constants, respectively ($C_{11}$, $C_{33}$ $C_{12}$, $C_{13}$, $C_{44}$ for the hexagonal symmetry, and additional $C_{22}$, $C_{23}$, and $C_{55}$ for the orthorhombic symmetry).
The polycrystalline Young's modulus, $E=9BG/(3B+G)$ was calculated using Hill's average of the bulk, $B$, and shear modulus, $G$ \cite{nye1985physical,hill1952elastic}.
The polycrystalline Poisson's ratio, $\nu$, and the directional Young's moduli, $E^{\langle 001\rangle}$, $E^{\langle 010\rangle}$, and $E^{\langle 100\rangle}$, were calculated following Ref.~\cite{nye1985physical}. 
Consequently, the out-of-plane and the in-plane Young's moduli, i.e., normal and parallel to the metal/ceramic layers, respectively, were calculated as 
\begin{align}
   E^{\perp}&=E^{\langle 001\rangle} \ , \\
   E^{\parallel}&=\frac{1}{2} ( E^{\langle 010\rangle} + E^{\langle 100\rangle} ) \ .
\end{align}


\section{Results and discussion}

\subsection{Stability trends}
The searched chemical and phase space of hypothetical MAB phases contains all combinations of the group 4--7 transition metals (M elements) with Al, Si, Ga, Ge or In (A elements) and 10 phase prototypes (Fig.~\ref{FIG: prototypes}) for each elemental combination. 
Though Tc-based MABs have expectably low appeal for applications\cite{wildung1979technetium}, they are included for completeness. 
Our first aim is predicting stability trends, preferential phase prototypes, and energetically-close competing MABs.

The energy of formation, $E_f$, serves as a basic chemical stability indicator, allowing to pre-select hypothetically (meta)stable MABs further tested for stability with respect to ``small'' elastic deformations and phonon vibrations, i.e., mechanical and dynamical stability.
For each (M, A) combination, we identify (i) the lowest-energy phase, and (ii) energetically-close phases, i.e., within an energy threshold, $E_f^{\text{thr}}$, from the lowest-energy phase (here $E_f^{\text{thr}}\coloneqq0.25$~eV/at.). 
For thereby selected MABs, we verify positive definiteness of the corresponding elastic constants matrix (mechanical stability condition\cite{mouhat2014necessary}) and the absence of imaginary phonon modes in the phonon spectra (dynamical stability condition).
Note that MABs passing these stability conditions may be still metastable against the decomposition to any competing binary or ternary non-MAB compounds~\cite{khazaei2019novel,carlsson2022theoretical} not considered in this work.

\begin{figure*}[h!t!]
    \centering
    \includegraphics[width=2\columnwidth]{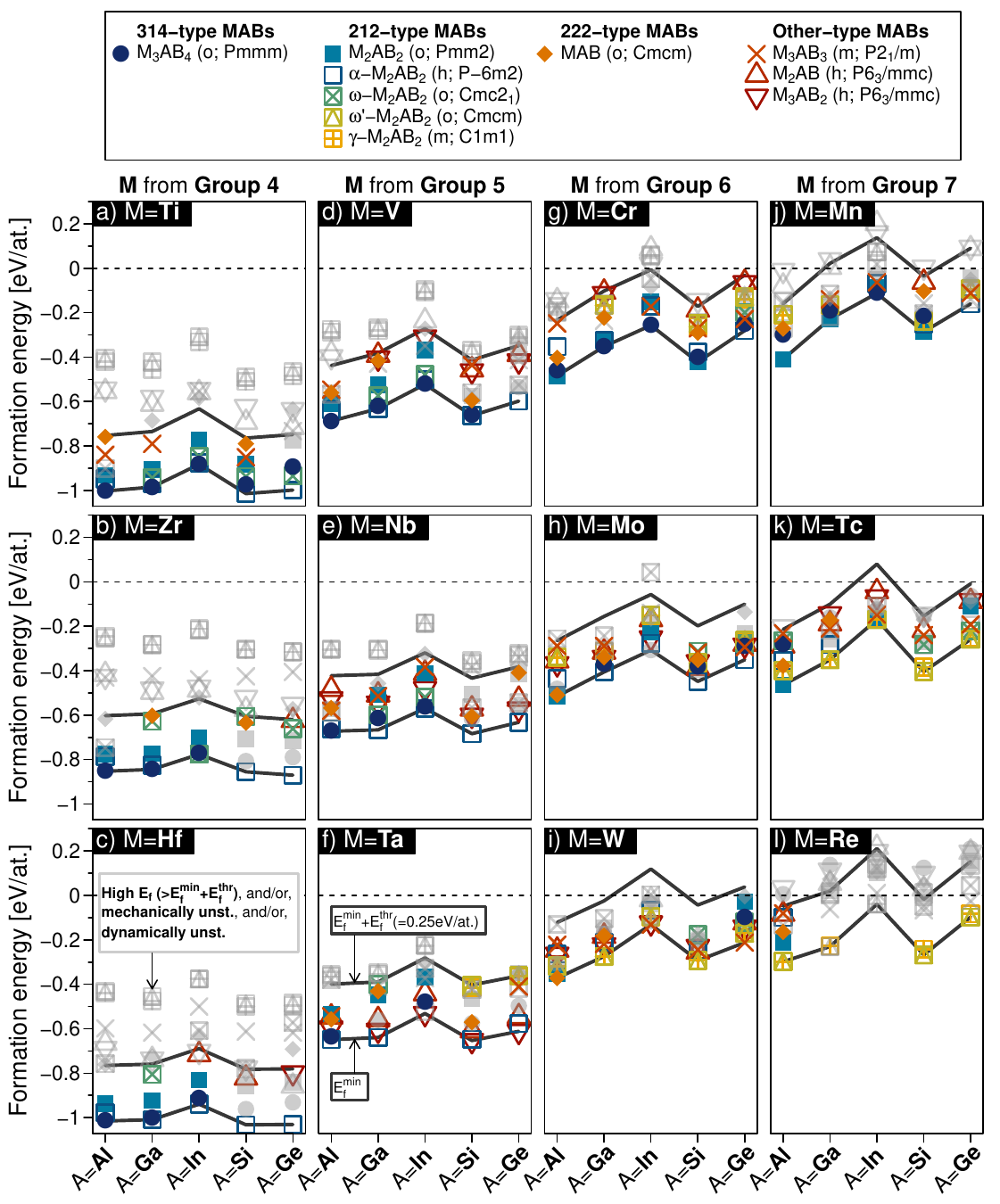}
    \caption{\small
    {\bf{Trends in the phase stability of MABs}}, as quantified by the formation energy, $E_f$ (see Eq.~\eqref{Eq: Ef}). 
    For each (M, A) combination, the black solid lines guide the eye for MABs energetically close to the lowest-energy phase, i.e. within the $E_f$ threshold, $E_f^{\text{thr}}=0.25$\;eV/at. 
    MABs that are above $E_f^{\text{thr}}$, and/or, mechanically, and/or dynamically unstable are depicted in grey.
    All MABs marked by colour are mechanically and dynamically stable.
    Trends in the corresponding per-atom volumes, densities, and atomic difference ratios are shown in Suppl. Fig.~\ref{SUPPL FIG: volume}, Suppl. Fig.~\ref{SUPPL FIG: density}, and Suppl. Fig.~\ref{SUPPL FIG: at diff ratio}, respectively. 
	}
\label{FIG: Ef}
\end{figure*}

The predicted $E_f$ trends are depicted in Fig.~\ref{FIG: Ef}.
Out of all hypothetical MABs (total 600), about 50\% (317 MABs) fulfil the above selection criteria (energetic, mechanical and dynamical stability), and are visualised by coloured symbols.
Together with the analysis of the corresponding structural properties---per-atom volumes, $V_{\text{per-at.}}$ (Suppl. Fig.~\ref{SUPPL FIG: volume}), and densities, $\rho$ (Suppl. Fig.~\ref{SUPPL FIG: density})---results in Fig.~\ref{FIG: Ef} lead to the following observations:

\begin{enumerate}
    \item[(3.1a)] {\bf{Trends in the energetic ($E_f$) stability are mainly driven by M}}. 
    Typically, the MABs' {\bf{formation energy increases for M from the group 4$\to$5$\to$6$\to$7}}. Exemplarily, $E_f$ of the M$_2$AlB$_2$ phase gradually  increases from $-0.93$ to $-0.41$~eV/at. for M changing as  Ti$\to$V$\to$Cr$\to$Mn.
    Generally, the $E_f$ increase is accompanied by {\bf{the volume}} ($V_{\text{per-at.}}$) {\bf{decrease}} and {\bf{the density}} ($\rho$) {\bf{increase}}.
    Changing the M's period (4$\to$5$\to$6) has a relatively minor effect on $E_f$, $V_{\text{per-at.}}$, and $\rho$. 
    
    \item[(3.1b)] {\bf{For a given M, Al-containing (In-containing) MABs typically exhibit the lowest (highest) $E_f$}} among the here-considered phase prototypes.
    {\bf{In-containing MABs also show the highest volume and the lowest density}}, possibly due to the large atomic radius of In (compared to Al, Ga, Si, and Ge), which is closer to that of M elements (see the atomic difference ratio analysis in Suppl. Fig.~\ref{SUPPL FIG: at diff ratio}).
    Most often, $E_f$ increases for A changing as Al$\to$Si$\to$Ga$\to$Ge$\to$In. 
    The choice of A more significantly impacts stability for MABs that contain M from groups 6--7 (compared with M from groups 4--5).
    There are no stable MABs combining Re and In. 

    \item[(3.1c)] {\bf{The M element's group}} also notably influences {\bf{which phase prototypes are energetically favourable}} and {\bf{how many energetically-close MABs exists}}.
    The M$_3$AB$_4$ prototype always yields the lowest volume and the highest density, while the M$_2$AB and M$_3$AB$_2$ are typically the least dense.
        
    \begin{itemize} 
        \item For M from the {\bf{group 4 ($=$ Ti, Zr, Hf)}}, the lowest-energy MABs are M$_3$AB$_4$ ($\text{A}=\{$Al, Ga, In\}) and $\alpha$-M$_2$AB$_2$ ($\text{A}=\{$Si, Ge\}), followed by M$_2$AB$_2$ and $\omega$-M$_2$AB$_2$. The Ti--Al--B and Ti--Si--B systems exhibit the highest number of possible MABs: M$_3$AB$_4$, $\alpha$-M$_2$AB$_2$, M$_2$AB$_2$, $\omega$-M$_2$AB$_2$, M$_3$AB$_3$, and MAB, where the last two exhibit the highest $E_f$, thus are the least likely to form.

        \item For M from the {\bf{group 5 ($=$ V, Nb, Ta)}}, the lowest-energy MABs are M$_3$AB$_4$, $\alpha$-M$_2$AB$_2$ or M$_3$AB$_2$, where the last one only concerns Ta--A--B systems with $\text{A}=\{$Si, Ge, In\}.
        Compared to M from the group 4, there are more competing MABs.
        The MAB phase prototype becomes a competing phase, particularly for $\text{A}=\{$Al, Si\}.
        The M$_2$AB$_2$ and $\omega$-M$_2$AB$_2$ prototypes are competing phases in the V--A--B and Nb--A--B systems for $\text{A}=\{$Ga, In\}.        

        \item For M from the {\bf{group 6 ($=$ Cr, Mo, W)}}, the lowest-energy MABs are M$_3$AB$_4$, $\alpha$-M$_2$AB$_2$, M$_2$AB$_2$, MAB (Mo--Al--B and W--Al--B systems), $\omega'$-M$_2$AB$_2$ (W--Si--B and W--Ga--B systems), and M$_3$AB$_3$ (W--Ge--B system). While nearly all the here-considered phase prototypes fall within the $E_f$ threshold, some are dynamically unstable. 

        \item For M from the {\bf{group 7  ($=$ Mn, Tc, Re)}}, the lowest-energy MABs are $\omega'$-M$_2$AB$_2$, M$_2$AB$_2$ (Mn--A--B system for $\text{A}$=\{Al, Si, Ga\}), $\alpha$-M$_2$AB$_2$ (Mn--Ge--B system), M$_3$AB$_4$ (Mn--In--B system).
        Mn--Al--B and Re--Al--B exhibit the highest number of competing MABs. 

    \end{itemize}
  
\end{enumerate}

\begin{figure*}[h!t!]
    \centering
    \includegraphics[width=2\columnwidth]{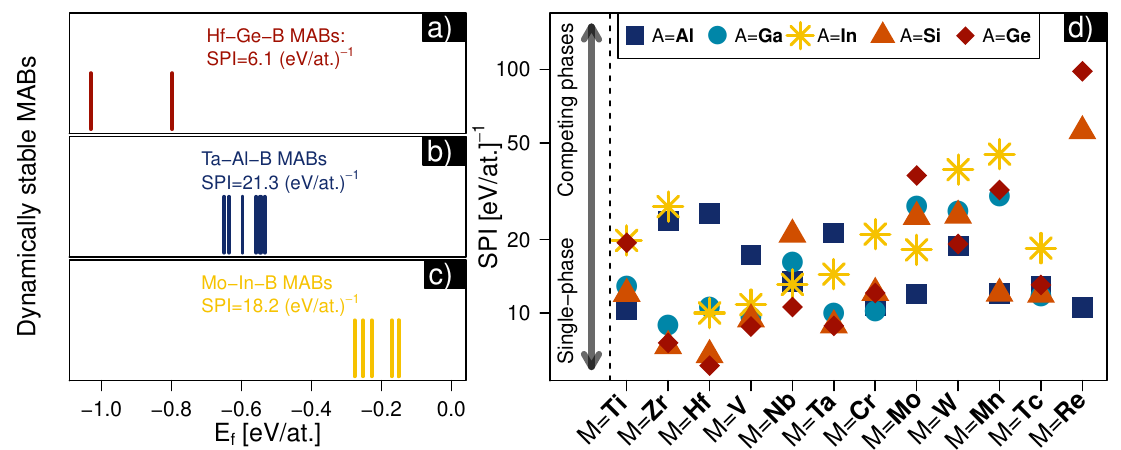}
    \caption{\small
    {\bf{Prediction of the MABs' propensity to form a single-phase compound}}, as quantified by the {\it{single-phase indicator}}, SPI, see Eq.~\eqref{Eq: SPI}.
    (a,b,c) Examples of formation energy ($E_f$) spectra used to derive SPI. The spectra contain all dynamically stable phases (identified in Fig.~\ref{FIG: Ef}) for the respective (M, A) combination.
    (d) The SPI descriptor for all (M, A) combinations. Low SPI values indicate tendency to form as single-phase MAB, whereas high SPIs are a sign of several energetically-close competing MABs. Mind the log scale of the $y$-axis (SPI). 
	}
\label{FIG: SPI}
\end{figure*}

Calculated with respect to energies of the constituting elements (i.e. their chemical potentials), absolute $E_f$ values in Fig.~\ref{FIG: Ef} may change under (highly) non-equilibrium synthesis conditions, reflected by changes of the reference chemical potentials (see a case-study of TaN\cite{stampfl2003metallic}).
Consequently, this may alter relative stability order of energetically-close MABs.
The actual $E_f$ distribution of all dynamically stable MABs, however, will be less affected by ``small'' variations of chemical potentials. 
In particular, a system in which the lowest-energy phase exhibits a ``large'' $E_f$ separation from competing phases will likely enable single-phase MAB formation (supposing it will not decompose into binary or ternary non-MAB compounds), contrarily to a system with many energetically-close phases.
To quantify this intuitive idea, we take inspiration in the EFA (entropy forming ability) descriptor by Curtarolo and co-workers\cite{sarker2018high,kaufmann2020discovery} and introduce a {\it{single-phase indicator}}, SPI.
Considering the $E_f$ distribution of competing MABs in a given M--A--B system, SPI is calculated as
\begin{equation}
    \text{SPI}(n)=\{\sigma[\text{spectrum}(E_f(n))] \}^{-1} \ , 
    \label{Eq: SPI}
\end{equation}
where $\sigma$ is a standard deviation of energies, $E_f$. 
A ``low'' SPI suggests a high propensity to form a single-phase MAB, similar to Curtarolo's ``low'' EFA pointing towards formation of a single-phase high-entropy ceramic\cite{sarker2018high}.
An example $E_f$ spectra together with the resulting SPI are given in Fig.~\ref{FIG: SPI}(a,b,c).
The SPIs evaluated for all elemental combinations (Fig.~\ref{FIG: SPI}d) render the following hypotheses:
\begin{itemize}
    \item Hf and Zr combined with Si or Ge are likely to form single-phase MABs ($\text{SPI}\approx$6--7\;(eV/at.)$^{-1}$), if not decomposed into other non-MAB compounds.
    
    \item Contrarily, Re in combination with Si ($\text{SPI}\approx{98}$\;(eV/at.)$^{-1}$) or Ge ($\text{SPI}\approx{56}$\;(eV/at.)$^{-1}$), and Mn and W in combination with In ($\text{SPI}$ of $\approx{37}$ and 45\;(eV/at.)$^{-1}$, respectively) do not provide a suitable basis for stabilisation of single-phase MABs.

    \item For A$=$Al---the most typical A element in experimentally reported MABs---there is no extremely high or low propensity to single-phase MABs formation. Examples of M elements yielding relatively ``lower'' and ``higher'' SPI are M$=\{$Ti, Cr, Mo, Mn, Tc, Re\}, with $\text{SPI}\approx$10--13\;(eV/at.)$^{-1}$, and M$=\{$Zr, Hf, Ta\} with $\approx$21--26\;(eV/at.)$^{-1}$, respectively. 
\end{itemize}

The SPI reflects energetic aspects only (here, zero Kelvin formation energies of dynamically stable MABs).
In practice, however, single-phase MAB formation will be driven also by kinetic factors and specific experimental setup (e.g. sputtering from a ternary vs. elemental targets).
A simple add-on to the SPI may be considering the volumetric proximity of competing phases based on the $V_{\text{per-at}}$ data in Suppl. Fig.~\ref{SUPPL FIG: volume}.

Supposing the MABs' layered character enables a relatively ``easy'' transformation pathway from one phase prototype to another, the availability of energetically-close phases (intermediate-to-high SPI) with similar volumes as the energetically most preferred phase may actually be beneficial: facilitating transformation toughening during volume-changing mechanical deformation.
For example, cubic Ti$_{0.5}$Al$_{0.5}$N subject to [001] tension undergoes local lattice transformations to the energetically-close wurtzite structure, thus improving toughness\cite{sangiovanni2020strength}.  
An additional important condition for the phase transformation may be that the M:A:B ratio remains unchanged. 
Our $E_f$ and volumetric analysis points towards Cr$_2$AlB$_2$, Re$_2$AlB$_2$, Cr$_2$SiB$_2$, W$_2$SiB$_2$, and Mn$_2$SiB$_2$ as to MAB phases with intermediate $\text{SPI}\approx$11--25\;(eV/at.)$^{-1}$, favouring the 2:1:2 stoichiomtery and exhibiting energetically-close 2:1:2-type phases with volumes by 1\%--10\% larger (smaller in case of Mn$_2$SiB$_2$), thus possibly forming under tensile (compressive) strains.

\subsection{Trends in the electronic structure}
Our stability predictions revealed in total 317 (meta)stable MABs (see coloured symbols in Fig.~\ref{Eq: Ef}).
Due to various crystal symmetries, chemistry, and elemental composition, however, understanding trends in their mechanical properties may be difficult. 
To shed light on their most fundamental similarities and differences, we first investigate their electronic density of states (DOS) near the Fermi level.

\begin{figure*}[h!t!]
    \centering
    \includegraphics[width=2\columnwidth]{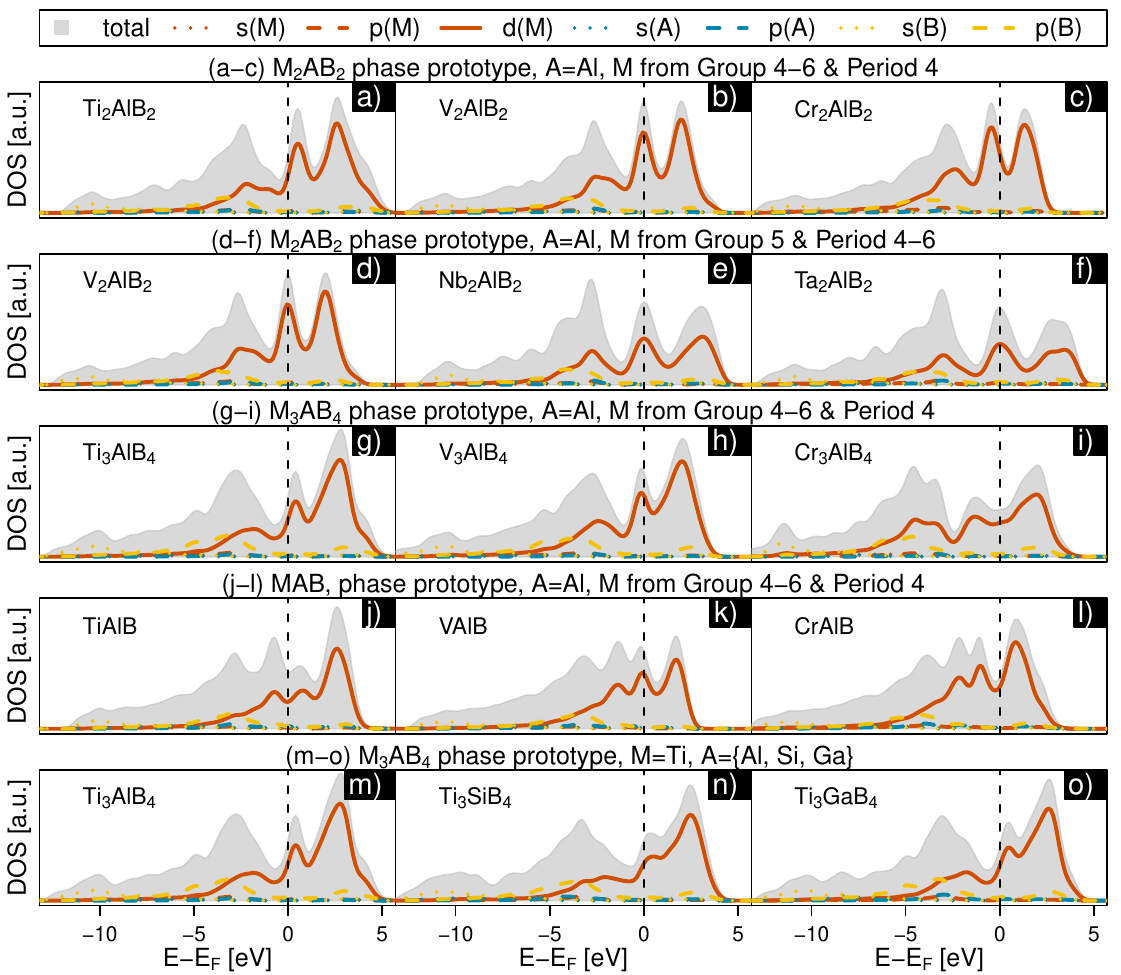}
    \caption{\small
     {\bf{Electronic density of states (DOS) for representative MABs illustrating general DOS shape depending on the phase prototype and elemental composition}}.
     The grey-shaded area denotes the total DOS, while the red, blue,and yellow lines are partial contributions from the M, A, and B element (dotted line: s-electrons, dashed line: p-electrons, solid line: d-electrons). 
     The zero energy always denotes the Fermi level ($E_F$).
     (a--c) {\bf{The role of the group from which M is chosen}}, exemplified by the {\bf{M$_2$AB$_2$}} phase prototype. The DOS shapes are very similar, while the $E_F$ shifts. 
     (d--f) {\bf{The role of the period from which M is chosen}}, exemplified by the {\bf{M$_2$AB$_2$}} phase prototype. Both the DOS shape and $E_F$ are very similar. 
     (g--i) {\bf{The role of the group from which M is chosen}}, exemplified by the {\bf{M$_3$AB$_4$}} phase prototype. The DOS shape is very similar (slightly differing for the eneregtically least stable Cr$_3$AlB$_4$), while the $E_F$ shifts. 
     (j--l) {\bf{The role of the group from which M is chosen}}, exemplified by the {\bf{MAB}} phase prototype. The DOS shape is very similar, while the $E_F$ shifts. 
     (m-l) {\bf{The role of the A element}}, exemplified by the {\bf{M$_3$AB$_4$}} phase prototype. The DOS shape changes slightly, while the $E_F$ position relative to the nearest DOS peak remains nearly constant. 
	}
\label{FIG: DOS}
\end{figure*}

Fig.~\ref{FIG: DOS} presents DOS of selected representative MABs, illustrating general trends observed also for other phase prototypes and (M, A) combinations. 
The energy range in focus is $\approx{[-15,5]}$~eV, where 0 corresponds to the Fermi level, $E_F$.
From a qualitative DOS analysis, we infer the following:
\begin{enumerate}
    \item[(3.2a)] The DOS of all MABs has metallic character and the Fermi level vicinity is dominated by the transition metal d-electrons (Fig.~\ref{FIG: DOS}a--o), suggesting that the M element significantly influences mechanical properties.  
    
    \item[(3.2b)] The most decisive factor for the general shape of DOS is the phase prototype, whereas the $E_F$ position with respect to the nearest peaks is mainly governed by the M element's group (Fig.~\ref{FIG: DOS}a--l). Exceptions (e.g. Cr$_3$B$_4$ in Fig.~\ref{FIG: DOS}i) may be rationalised by high formation energy of the corresponding phase prototype for given M.
    
    \item[(3.2c)] For the same phase prototype and M from the same group in the periodic table (i.e. with the same valence electron concentration), both the shape of DOS and $E_F$ are typically nearly the same (Fig.~\ref{FIG: DOS}d--f).

    \item[(3.2d)] Changing the A element influences the general shape of DOS, however, to a lesser extent than changing the phase prototype (Fig.~\ref{FIG: DOS}m--o). The relative $E_F$ position with respect to the neighbouring peaks is nearly uninfluenced.

\end{enumerate}

Our results indicate that the main features of DOS near $E_F$ are dictated by the phase prototype, which may be the most natural mean of sorting MABs when searching for trends in mechanical properties.
The group of the M element---in other words, the number of M's valence electrons---influences the Fermi level position with respect to the closest DOS minimum or peak, thus, may be a crucial factor for optimisation of mechanical properties.
Shifting the $E_F$ for different phase prototypes, however, will likely lead to filling different states and an in-depth analysis (out of our scope) would be necessary to understand how these impact the MABs' mechanical response.

\subsection{Stiffness and ductility indicators}
This section focuses on predicting trends in mechanical properties of MAB phases via phenomenological elastic-constants-based descriptors. 
Specifically, Young's modulus ($E$), shear-to-bulk modulus ratio ($G/B$), and Cauchy pressure ($CP$) are used to compare all dynamically stable MAB candidates (all coloured symbols in Fig.~\ref{Eq: Ef}, comprising 317 MABs) in terms of theoretical stiffness and ductility.
Although ductility is a complex property---dictated by structure, density, and mobility of extended crystallographic defects over different length and time scales---$C_{ij}$-based indicators have served as common trend-givers in the family of refractory ceramics~\cite{kindlund2019review,sangiovanni2010electronic,balasubramanian2018valence,koutna2021high,moraes2018ab,kretschmer2022high} with reasonably similar crystal and electronic structures.

Due to the non-cubic crystal symmetry and layered architecture, MAB phases should generally exhibit an anisotropic elastic response.
The degree of anisotropy can be estimated by the universal anisotropy index, $A^{\text{U}}$\cite{ranganathan2008universal}, identically zero for locally isotropic single crystals.
In our case, $A^{\text{U}}$ varies between 0.01 ($\alpha$-Nb$_2$InB$_2$) and 6.35 (Ta$_3$InB$_4$), where nearly 80\% of all MABs exhibit $A^U\leq0.5$ (Suppl. Fig.~\ref{SUPPL FIG: Anisotropy}).
The most energetically preferred phase for a given (M, A) combination yields $A^U=0.01$--1.43, comparable to $A^{\text{U}}=0.01$--1.88 of transition metal diboride ceramics (MB$_2$, M from the group 4--7; $A^U$ was evaluated based on  data from Ref.~\cite{leiner2023energetics}). 
Interestingly, $A^U$ of the $\alpha$-M$_2$AB$_2$ phase prototype is fairly independent of elemental composition ($A^U=0.11\pm0.14$), while others, e.g., the M$_2$AB$_2$ and M$_3$AB$_4$ phase prototype, exhibit larger $A^U$ variations ($A^U=0.54\pm0.49$ and $A^U=0.66\pm1.16$, respectively).

\begin{figure*}[h!t!]
    \centering
    \includegraphics[width=2\columnwidth]{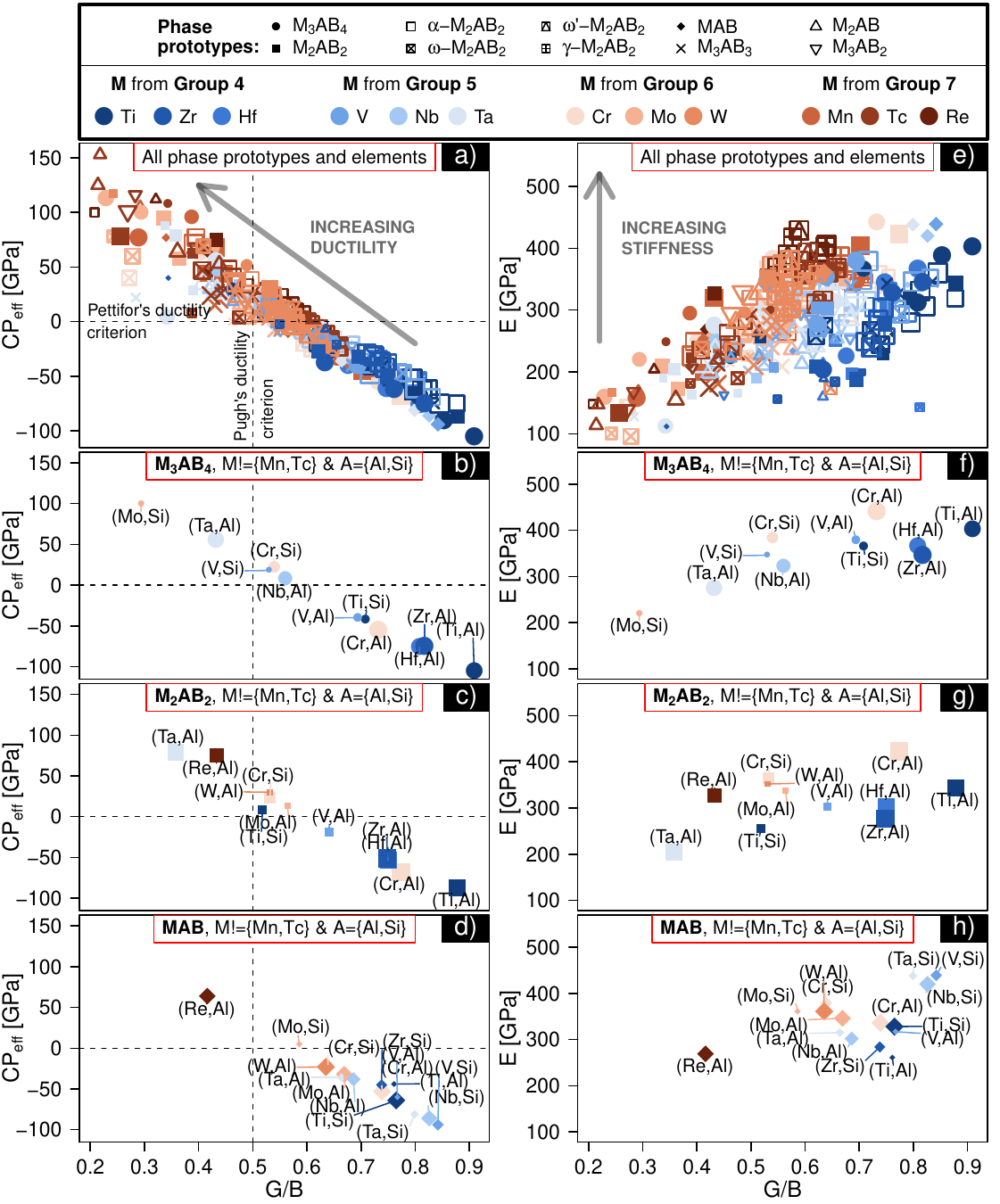}
    \caption{\small
    {\bf{Trends in theoretical ductility (a--d) and stiffness (e--h) of MAB phases estimated via elastic-constants-based descriptors}}: effective Cauchy pressure ($CP_{\text{eff}}$, Eq.~\ref{Eq: CPeff}), polycrystalline shear-to-bulk modulus ratio ($G/B$), and polycrystalline Young's modulus ($E$).    
    The dashed lines in (a--d) guide the eye for Pettifor's\cite{pettifor1992theoretical} and Pugh's\cite{pugh1954xcii} semi-empirical ductility criteria (commonly used for ceramics\cite{kindlund2019review,sangiovanni2010electronic,balasubramanian2018valence,koutna2021high,moraes2018ab,kretschmer2022high} but originally developed for metals, these criteria should be treated only on a qualitative level).
    Panels (a) and (e) show results for all stable phases (marked by colour in Fig.~\ref{FIG: Ef}), while panels (b--c) and (f--h) focus on the most common phase prototypes and elements (excluding Mn, Tc, Ga, Ge, In). 
    The underlying (phase-, M-element-, and A-element-resolved) $B$, $G$, and $E$ values for all MABs are shown in Suppl. Fig.~\ref{SUPPL FIG: B}, Suppl. Fig.~\ref{SUPPL FIG: G}, and Suppl. Fig.~\ref{SUPPL FIG: E}, respectively. The corresponding Poisson's ratios are shown in Suppl. Fig.~\ref{SUPPL FIG: nu}.  
	}
\label{FIG: Ductility indicators}
\end{figure*}

In the first step (Fig.~\ref{FIG: Ductility indicators}), we disregard the MABs' elastic anisotropy and evaluate ``effective'' strength and ductility indicators using the polycrystalline moduli ($B$, $G$, $E$) and averaging the directional Cauchy pressure values,
\begin{align}
    CP_{\text{eff}}&=CP^{\perp} + CP_{[1]}^{\parallel} + CP_{[2]}^{\parallel} = \nonumber \\ 
   &=\frac{C_{12}-C_{66}}{3} + \frac{C_{13}-C_{44}}{3} + \frac{C_{23}-C_{55}}{3},  
\label{Eq: CPeff}   
\end{align}
where $CP^{\perp}$ ($CP^{\parallel}_{[1]}$, $CP^{\parallel}_{[2]}$) are out-of-plane (in-plane) Cauchy pressure, i.e. orthogonal and parallel to the M--B/A layers. 
The obtained ductility map ($CP_{\text{eff}}$ vs. $G/B$ in Fig.~\ref{FIG: Ductility indicators}a) indicates an important role of the M element. 
The strongest trend observed is a ductility increase for M from the group 4$\to$5$\to$6$\to$7.  
Among the most ductile MABs containing common elements (excluding Ga, Ge, In, Mn, Tc) are V$_2$SiB, Nb$_2$SiB, Ta$_2$AlB$_2$, Ta$_3$AlB$_4$, Ta$_2$SiB, Cr$_3$AlB$_3$, Mo$_2$AlB$_2$, Mo$_2$SiB$_2$, Mo$_3$AlB$_3$, Mo$_3$SiB$_3$, W$_3$SiB$_2$, Re$_2$AlB$_2$, and ReAlB.
There are, however, several outliers.
For example, Cr$_2$AlB$_2$ and MoAlB (M from the group 6) are predicted to be surprisingly brittle ($CP_{\text{eff}}<-30$~GPa), while Ti$_3$GeB$_4$ and ZrGaB (M from the group 4) would be surprisingly ductile ($CP_{\text{eff}}>30$~GPa).
These outliers may be explained by (i) energetic reasons ($E_f$ ``high'' above that of the lowest-energy phase, e.g., ZrGaB with $E_f$ 0.24~eV/at above $E_f$ of the most stable Zr$_3$GaB$_4$), by (ii) differences between the phase prototypes, i.e. different optimal position of the Fermi level (inducing more ductile/brittle response to deformation), by (iii) effects of the A element (recall that In-containing MABs were always the least energetically stable and the least dense), or by (iv) elastic anisotropy.
To support the energetic argument (i), all data points in Fig.~\ref{FIG: Ductility indicators} are scaled based on their $E_f$ difference from the lowest-energy phase for a given (M, A) combination, so that smaller symbol sizes correspond to larger $E_f$ differences.

The so far experimentally reported MABs most often crystallised in the  M$_3$AB$_4$, M$_2$AB$_2$, and MAB type phase, depicted in Fig.~\ref{FIG: Ductility indicators}b--d.
Here, Ta$_3$AlB$_4$, Ta$_2$AlB$_2$, Re$_2$AlB$_2$, and ReAlB stand out in terms of theoretical ductility.

Fig.~\ref{FIG: Ductility indicators}e--h shows the relationship between the elastic stiffness and ductility, estimated by the polycrystalline Young's modulus, $E$, and the shear-to-bulk modulus ratio, $G/B$, respectively.
To prevent failure during mechanical loads, one seeks a compromise between high $E$ and low $G/B$, providing an atomic-level basis for initially hard but then reasonably plastic response to deformation. 
For the here-studied MABs, $E$ varies significantly---between 96~GPa ($\omega$-Mo$_2$GeB$_2$) and 441~GPa (Cr$_3$AlB$_4$)---and seems to be less controlled by the M element than ductility.

Suggested already by low density of In-containing MABs (Section 3.1, Suppl. Fig.~\ref{SUPPL FIG: density}), their $E$ moduli are generally low, 232$\pm$50~GPa, where the standard deviation represents values from various M elements and phase prototypes.
Al- and Si-containing MABs, in contrast, posses relatively high $E$ values, 312$\pm$60~GPa.
There is eleven MABs with $E>400$~GPa. 
Only one contains a group 4 M element (Ti$_3$AlB$_4$) and the top three are Cr$_3$AlB$_4$, TaSiB, and VSiB possessing $G/B$ of 0.73, 0.80 0.84, respectively, thus illustrating the typical inverse relationship between stiffness and ductility. 
A combination of relatively high Young's modulus ($E>350$~GPa) and low $G/B$ ratio ($G/B<0.55$) is shown by Cr$_2$SiB$_2$, Cr$_3$SiB$_4$, W$_2$AlB$_2$, $\alpha$-W$_2$SiB$_2$, and $\gamma$-Re$_2$GeB$_2$.
Among them, Cr$_3$SiB$_4$ is the lowest-energy phase in the Cr--Al--B system, and the other ones are energetically close to their most favourable phases ($\Delta E_f=0.01$--0.05~eV/at.).

\begin{figure*}[h!t!]
    \centering
    \includegraphics[width=2\columnwidth]{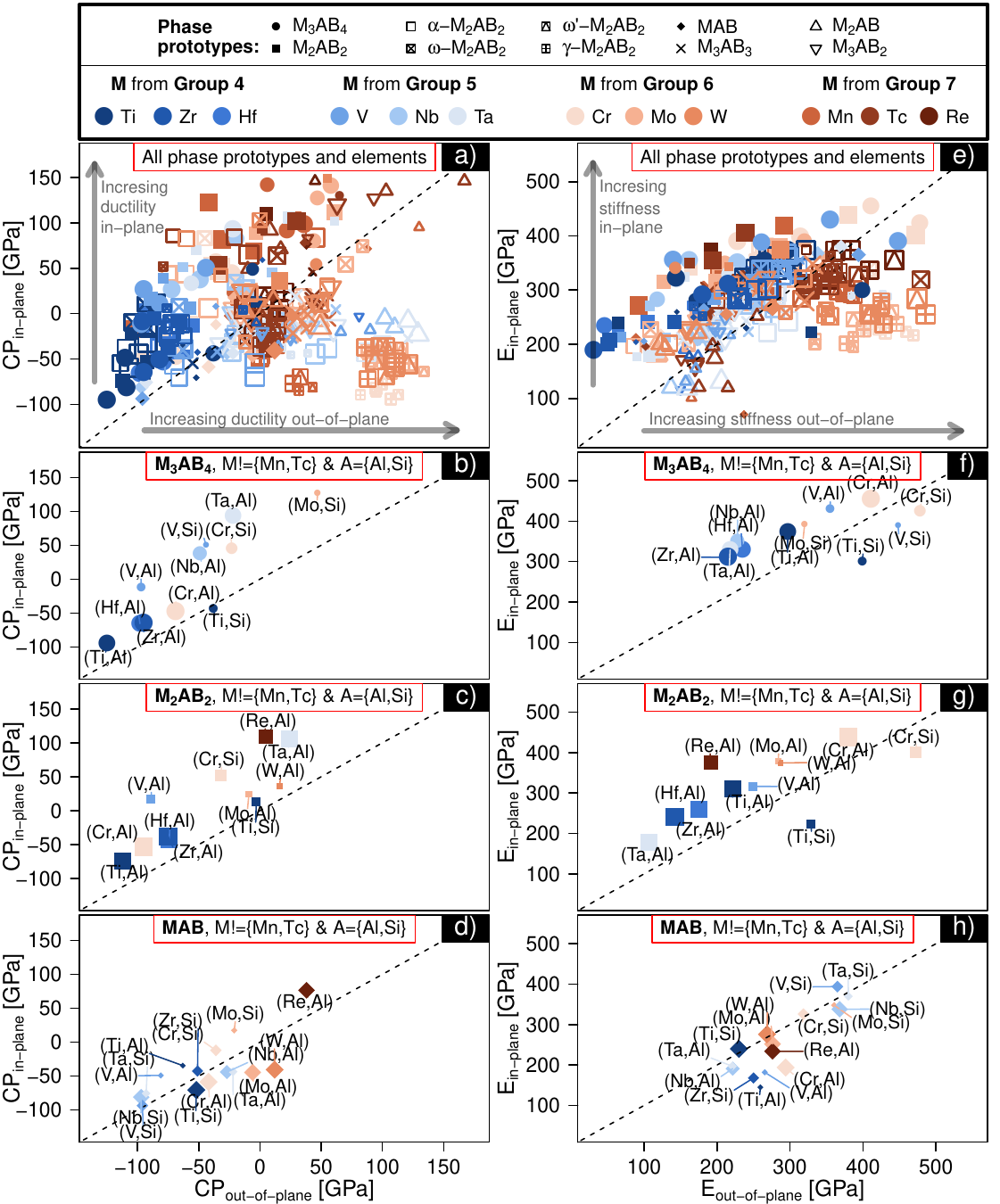}
    \caption{\small
      {\bf{Trends in theoretical in-plane vs. out-of-plane ductility (a--d) and stiffness (e--h) of MAB phases estimated via elastic-constants-based descriptors}}: directional Cauchy pressure ($CP_{\text{in-plane}}=CP^{\parallel}$, $CP_{\text{out-of-plane}}=CP^{\perp}$) and directional Young's modulus ($E_{\text{in-plane}}=E^{\parallel}$, $E_{\text{out-of-plane}}=E^{\perp}$).   
      The dashed diagonal lines guide the eye for the case of equal in-plane and out-of-plane values. 
      Panels (a) and (e) show results for all stable phases (marked by colour in Fig.~\ref{FIG: Ef}), while panels (b--c) and (f--h) focus on the most common phase prototypes and elements (excluding Mn, Tc, Ga, Ge, In). 
	}
\label{FIG: Directional E and CP}
\end{figure*}

As noted at the beginning of this section, elastic response of MABs is strongly directional. 
This is illustrated in Fig.~\ref{FIG: Directional E and CP} presenting two Cauchy pressure and Young's modulus values: in-plane, i.e., parallel to the metal/ceramic layers (denoted by $CP^{\parallel}$, $E^{\parallel}$), and out-of-plane, i.e. orthogonal to the metal/ceramic layers (denoted by $CP^{\perp}$, $E^{\perp}$). 
The latter is aligned with the most typical growth direction. 
According to Fig.~\ref{FIG: Directional E and CP}a, about 20\% of all MABs can be seen as strongly Cauchy-pressure anisotropic, with $|CP^{\parallel}-CP^{\perp}|>100$\;GPa.
Interestingly, the ratio of MABs with $CP^{\parallel}>CP^{\perp}$ and $CP^{\parallel}<CP^{\perp}$ is nearly 1:1.
Examples of extreme cases include Mn$_2$SiB$_2$ and $\alpha$-Re$_2$AlB$_2$ ($CP^{\parallel}-CP^{\perp}>150$~GPa, indicating superior in-plane ductility), or Ta$_2$SiB and $\gamma$-Mo$_2$GaB$_2$ ($CP^{\perp}-CP^{\parallel}>150$~GPa, indicating superior out-of-plane ductility).
Furthermore, the data show that (i) almost all MABs containing group 4 transition metals exhibit $CP^{\parallel}<CP^{\perp}$, and (ii) the $\omega'$-M$_2$AB$_2$, $\gamma$-M$_2$AB$_2$ and M$_2$AX prototypes show $CP^{\parallel}\ll CP^{\perp}$ for the group 6 M elements. 

Focusing on the most common phase prototypes and (M, A) combinations, the M$_3$AB$_4$ (Fig.~\ref{FIG: Directional E and CP}b) and the M$_2$AB$_2$ (Fig.~\ref{FIG: Directional E and CP}c) type phases are predicted to be more ductile in-plane, with extreme cases ($CP^{\parallel}\gg CP^{\perp}$) being Nb$_3$AlB$_4$, Ta$_3$AlB$_4$, Ta$_2$AlB$_2$, and Re$_2$AlB$_2$.
The MAB phase prototype (Fig.~\ref{FIG: Directional E and CP}d) can be seen as rather Cauchy-pressure isotropic.

Concerning Young's moduli, their in-plane and out-of-plane values Fig.~\ref{FIG: Directional E and CP}e--h do not show a simply reversed trend with respect to the directional Cauchy pressures. 
Besides Ti$_3$SiB$_4$, Ti$_2$SiB$_2$, and ZrAlB, all MABs containing group 4 transition metals and most MABs containing group 5 transition metals are stiffer in-plane compared to their [001] direction. 
This is intuitively expected due to relatively weak bonding between the ceramic (M--B) and metallic (A) layers. 
Similar to Fig.~\ref{FIG: Directional E and CP}a, the $\omega'$-M$_2$AB$_2$, $\gamma$-M$_2$AB$_2$ and M$_2$AX prototypes tend to cluster for the group 6 M elements and can be seen as outliers with significantly higher out-of-plane stiffness ($E^{\perp}-E^{\parallel}>150$~GPa).

Among the most common phase prototypes and (M, A) combinations, the M$_3$AB$_4$ (Fig.~\ref{FIG: Directional E and CP}f) and the M$_2$AB$_2$ (Fig.~\ref{FIG: Directional E and CP}g) type phase typically exhibit $E^{\parallel}>E^{\perp}$.
In particular, Nb$_3$AB$_4$, Ta$_3$AB$_4$, and Re$_2$AlB$_2$ yield $E^{\parallel}-E^{\perp}>100$~GPa.
The MAB phase prototype (Fig.~\ref{FIG: Directional E and CP}h) shows small differences between in-plane and out-of-plane Young's moduli, with the most anisotropic MABs being CrAlB, ZrSiB, and the energetically rather unlikely (smaller symbol sizes) TiAlB and VAlB.

\FloatBarrier

\subsection{Suggestions for the most promising MABs}
\begin{figure*}[h!t!]
    \centering
    \includegraphics[width=2\columnwidth]{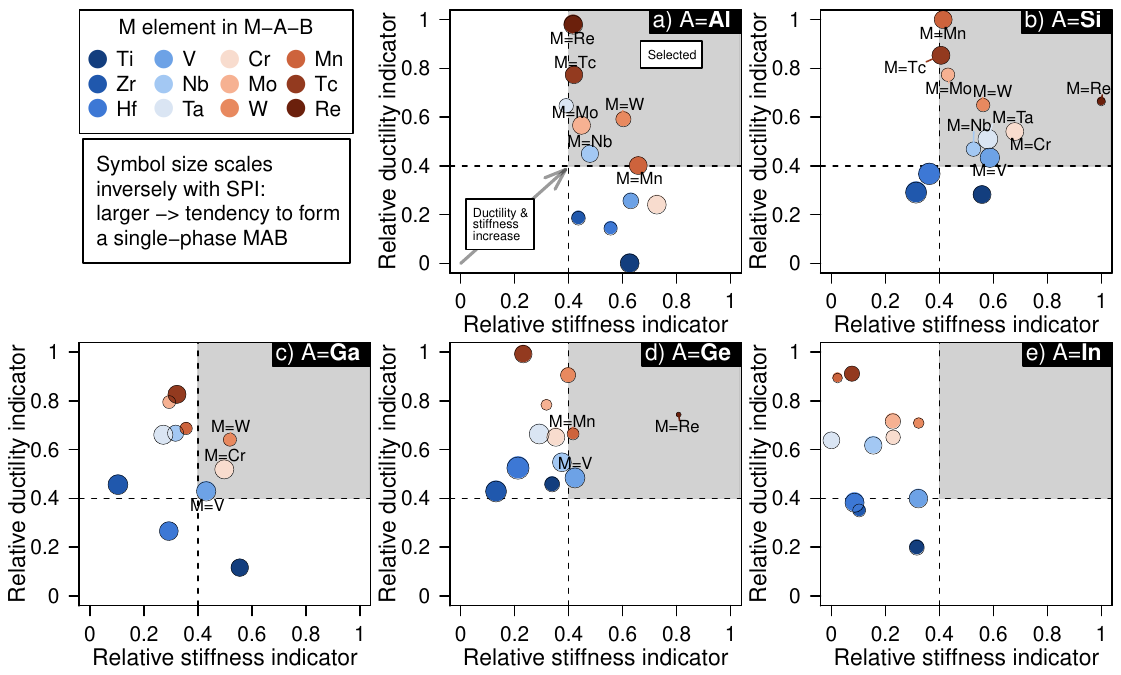}
    \caption{\small
      {\bf{Suggestion of suitable M--A--B systems for the development of MAB phases with favourable combination of stiffness and ductility.}}
      The relative stiffness (ductility) indicator is calculated as a weighted average of the polycrystalline Young's moduli, $E$, (effective Cauchy pressures, $CP_{\text{eff}}$) of all stable MABs in the respective M--A--B system (identified in Fig.~\ref{FIG: Ef}). 
      The weights are based on the formation energy difference from the lowest-energy phase---having a weight of 1---and the values are then normalised with respect to the global maxima (the Mn--Si--B and Re--Si--B system for ductility and stiffness, respectively).
      The symbol size scales with the {\it{single-phase indicator}} (SPI, Eq.~\eqref{Eq: SPI}): systems that tend to form single-phase MABs are depicted by larger symbols, while systems with energetically-close competing phases are depicted by smaller symbols.
      The dashed lines guide guide the eye for top 60\%-ranking material systems (top-right corner).
      }
\label{FIG: material selection}
\end{figure*}

In Fig.~\ref{FIG: material selection} we present M--A--B systems suitable for the development of MAB phases with favourable combination of stiffness and ductility.
These are identified based on Cauchy pressure and Young's modulus values of all mechanically and dynamically stable MABs, weighted according to their formation energy difference from the energetically most stable phase.

\begin{table*}[h!t!]
\caption{\small
    {\bf{Properties of the most energetically favourable MABs for promising (M, A) combinations}} (as identified in Fig.~\ref{FIG: material selection}). 
    The columns ``M'', ``A'', ``Phase'', $\Delta E_f$, and $\Delta V$ specify the elements, M and A, the phase prototype, the formation energy and volume difference from the most stable phase in the respective M--A--B system (phases with $\Delta E_f<0.05$~eV/at. are shown).
    Furthermore, the universal anistotropy index ($A^{\text{U}}$\cite{ranganathan2008universal}), the polycrystalline bulk, shear, and Young's modulus ($B$, $G$, $E$), the in-plane and out-of-plane Young's modulus ($E^{\parallel}$, $E^{\perp}$), the effective Cauchy pressure ($CP_{\text{eff}}$), the in-plane and out-of-plane Cauchy pressure ($CP^{\parallel}$, $CP^{\perp}$), the $B/G$ ratio, and the Poisson's ratio ($\nu$) are presented.
}
\centering
\footnotesize
\begin{tabular}{cccccccccccccccc}
\hline
\hline
M&A&Phase&$\Delta E_f$ & $\Delta V$&$A^{\text{U}}$&$B$&$G$&$E$&$E^{\parallel}$&$E^{\perp}$&$CP_{\text{eff}}$&$CP^{\parallel}$ & $CP^{\perp}$ & $G/B$ & $\nu$\\
& & & [eV/at.] & [\%] & & [GPa] & [GPa] & [GPa] & [GPa] & [GPa] & [GPa] & [GPa] & [GPa] & &  \\
\hline
{\bf{Nb}}&Al&M$_3$AB$_4$&0&0&0.21&227&127&322&351&228&8&37&$-49$&0.56&0.26\\
&Al&$\alpha$-M$_2$AB$_2$ &0.011&$-15.6$&0.03&182&130&315&297&268&$-36$&$-42$&$-23$&0.71&0.21\\
{\bf{Mo}}&Al&M$_2$AB$_2$&0&0&0.06&237&134&338&379&284&13&25&$-9$&0.56&0.26\\
&Al&MAB&0.006&$-4.6$&0.16&211&141&346&276&252&$-32$&$-45$&$-6$&0.67&0.23\\
{\bf{W}}&Al&MAB&0&0&0.18&229&146&361&269&276&$-23$&$-41$&12&0.64&0.24\\
&Al&M$_2$AB$_2$&0.02&4&0.06&260&138&352&375&287&30&36&16&0.53&0.27\\
{\bf{Mn}}&Al&M$_2$AB$_2$&0&0&0.09&237&166&404&419&292&$-45$&$-17$&$-102$&0.70&0.22\\
{\bf{Re}}&Al&$\gamma$-M$_2$AB$_2$&0&0&0.02&255&162&402&373&371&$-17$&$-23$&$-5$&0.64&0.24\\
&Al&$\omega'$-M$_2$AB$_2$&0.002&0&0.02&253&159&394&373&362&$-12$&$-19$&1&0.63&0.24\\
\hline
{\bf{V}}&Si&$\alpha$-M$_2$AB$_2$&0&0&0.09&218&156&379&338&310&$-48$&$-70$&$-4$&0.72&0.21\\
&Si&M$_3$AB$_4$&0.003&12.1&0.26&256&136&346&390&448&18&50&$-44$&0.53&0.28\\
{\bf{Nb}}&Si&$\alpha$-M$_2$AB$_2$&0&0&0.23&223&136&339&268&279&$-15$&$-45$&45&0.61&0.25\\
{\bf{Ta}}&Si&M$_3$AX$_2$&0&0&0.36&217&115&294&216&211&10&$-15$&60&0.53&0.27\\
&Si&$\alpha$-M$_2$AB$_2$&0.006&12&0.36&239&133&336&243&297&0&$-43$&87&0.56&0.27\\
&Si&M$_2$AX&0.044&$-5.5$&0.92&205&89&233&131&206&27&$-23$&129&0.43&0.31\\
{\bf{Cr}}&Si&M$_2$AB$_2$&0&0&0.15&269&143&365&402&472&23&$52$&$-32$&0.53&0.27\\
&Si&M$_3$AB$_4$&0.022&6&0.13&279&151&383&424&478&22&45&$-23$&0.54&0.27\\
&Si&$\alpha$-M$_2$AB$_2$&0.045&$-7.1$&0.11&237&152&375&351&256&$-22$&$-27$&$-14$&0.64&0.24\\
{\bf{Mo}}&Si&$\alpha$-M$_2$AB$_2$&0&0&0.09&253&147&369&324&296&4&$-8$&27&0.58&0.26\\
{\bf{W}}&Si&$\omega'$-M$_2$AB$_2$&0&0&0.32&264&151&381&284&474&$-3$&$-51$&94&0.57&0.26\\
&Si&$\gamma$-M$_2$AB$_2$&0.001&0&0.37&264&147&373&255&486&1&$-54$&111&0.56&0.26\\
&Si&M$_2$AX&0.038&$-13.3$&0.62&247&128&328&223&171&6&$-15$&49&0.52&0.28\\
&Si&$\alpha$-M$_2$AB$_2$&0.045&3.4&0.09&273&147&373&321&318&24&11&49&0.54&0.27\\
&Si&M$_3$AB$_3$&0.049&$-5.1$&0.25&213&114&290&304&259&12&$-1$&39&0.54&0.27\\
{\bf{Mn}}&Si&M$_2$AB$_2$&0&0&0.26&282&122&320&369&281&68&123&$-42$&0.43&0.31\\
&Si&$\alpha$-M$_2$AB$_2$&0.033&$-8.7$&0.11&237&94&249&292&250&77&102&27&0.40&0.32\\
&Si&$\gamma$-M$_2$AB$_2$&0.045&$-9.6$&0.14&237&158&388&306&366&$-32$&$-49$&0&0.67&0.23\\
&Si&$\omega'$-M$_2$AB$_2$&0.048&$-9.6$&0.2&236&148&366&284&316&$-19$&$-35$&13&0.63&0.24\\
\hline
{\bf{V}}&Ga&$\alpha$-M$_2$AB$_2$&0&0&0.02&183&146&346&367&295&$-61$&$-56$&$-70$&0.8&0.18\\
&Ga&M$_3$AB$_4$&0.011&13.5&0.14&223&143&353&387&261&$-20$&10&$-82$&0.64&0.24\\
{\bf{Cr}}&Ga&M$_3$AB$_4$&0&0&0.09&243&132&335&398&314&20&42&$-22$&0.54&0.27\\
&Ga&$\alpha$-M$_2$AB$_2$&0.028&$-14.9$&0.05&202&148&357&381&265&$-47$&$-33$&$-74$&0.73&0.2\\
&Ga&M$_2$AB$_2$&0.032&$-9$&0.15&224&119&304&372&268&20&47&$-34$&0.53&0.27\\
{\bf{W}}&Ga&$\omega'$-M$_2$AB$_2$&0&0&0.24&233&131&331&255&410&3&$-56$&121&0.56&0.26\\
&Ga&$\gamma$-M$_2$AB$_2$&0&0.1&0.23&234&135&339&255&411&$-2$&$-59$&110&0.58&0.26\\
\hline
{\bf{V}}&Ge&$\alpha$-M$_2$AB$_2$&0&0&0.05&204&144&349&328&289&$-38$&$-55$&-4&0.70&0.22\\
{\bf{Mn}}&Ge&$\alpha$-M$_2$AB$_2$&0&0&0.17&218&85&226&283&259&71&84&45&0.39&0.33\\
&Ge&M$_3$AB$_3$&0.048&$-1.7$&0.46&169&89&226&301&192&8&$-3$&29&0.53&0.28\\
\hline
\hline
\end{tabular}
\label{Tab: best}
\end{table*}

Consequently, the following elemental combinations are proposed:
\begin{itemize}
    \item {\bf{$\text{A}=\text{Al}$ and $\text{M}=\{$Nb, Mo, W, Mn, (Tc), Re\}}} (Fig.~\ref{FIG: material selection}a). The most energetically stable compounds---listed in Tab.~\ref{Tab: best}---are Nb$_3$AlB$_4$, Mo$_2$AlB$_2$, MoAlB, W$_2$AlB$_2$, WAlB, Mn$_2$AlB$_2$, $\gamma$-Re$_2$AlB$_2$, and $\omega'$-Re$_2$AlB$_2$ (Tab.~\ref{Tab: best}). Bulk WAlB\cite{zhang2023experimental,roy2023low}, bulk Mn$_2$AlB$_2$\cite{kota2018synthesis,roy2023low}, and thin film and bulk MoAlB\cite{achenbach2019synthesis,evertz2021low,sahu2022defects,zhang2023experimental,chen2019compressive} have already been synthesised. The experimental Young's modulus of MoAlB\cite{evertz2021low}, $E=379\pm 30$~GPa, compares well with our DFT value, $E=346$~GPa (Tab.~\ref{Tab: best}). 
    Furthermore, the here-predicted Mo$_2$AlB$_2$ was observed in HRTEM after topochemical deintercalation of Al from the single crystalline MoAlB\cite{alameda2018topochemical}.  

    With both Cauchy pressures positive ($CP^{\parallel}\approx{36}$~GPa, $CP^{\perp}\approx{16}$~GPa) and the lowest $G/B$ ($\approx{0.53}$), W$_2$AlB$_2$ stands out in terms of theoretical ductility. Mn- and Re-containing MABs exhibit the highest elastic stiffness. Mn$_2$AlB$_2$ is significantly stiffer in-plane, $E^{\parallel}\approx{419}$~GPa, $E^{\perp}\approx{292}$~GPa (comparable to room-temperature experimental value of 243\;GPa\cite{kota2018synthesis}), whereas  $\gamma$-Re$_2$AlB$_2$ is notably more isotropic, $E^{\parallel}\approx{373}$~GPa, $E^{\perp}\approx{371}$~GPa).
    
    \item {\bf{$\text{A}=\text{Si}$ and $\text{M}=\{$V, Nb, Ta, Cr, Mo, W, Mn, (Tc), (Re)\}}} (Fig.~\ref{FIG: material selection}b).
    The most energetically stable compounds are $\alpha$-V$_2$SiB$_2$, V$_3$SiB$_4$, $\alpha$-Nb$_2$SiB$_2$, Ta$_3$SiB$_2$, $\alpha$-Ta$_2$SiB$_2$, Ta$_2$SiB, Cr$_2$SiB$_2$, Cr$_3$SiB$_4$, $\alpha$-Cr$_2$SiB$_2$, $\alpha$-Mo$_2$SiB$_2$, $\omega'$-W$_2$SiB$_2$, $\gamma$-W$_2$SiB$_2$, W$_2$SiB, $\alpha$-W$_2$SiB$_2$, W$_3$SiB$_3$, Mn$_2$SiB$_2$, $\alpha$-Mn$_2$SiB$_2$, $\gamma$-Mn$_2$SiB$_2$, and $\omega'$-Mn$_2$SiB$_2$ (Tab.~\ref{Tab: best}). Re exhibits many competing phases. 

    Contrarily to Al-containing MABs, their Si counterparts are experimentally mostly unexplored. Considering outstanding oxidation properties of Si-alloyed boride ceramics\cite{aschauer2021ultra,kiryukhantsev2020mechanical,glechner2022influence}, we envision that also Si-containing MABs will attract attention.     
    
    As suggested by Cauchy pressure and Poisson's ratio values in Tab.~\ref{Tab: best}, Si-based MABs are slightly less brittle than Al-based MABs. With the 2:1:2 M:A:B chemistry, W exhibits three energetically- and volumetrically-close phase prototypes ($\Delta E_f=0.001$--0.045~eV/at., $\Delta V=0$--3.4\%) providing a basis to optimise mechanical properties. Similarly, Mn offers four energetically-close phases of the 2:1:2 chemistry. The $\alpha$-Mn$_2$SiB$_2$ exhibits the highest ductility indicators among all Si-containing MABs ($CP^{\parallel}\approx{102}$~GPa, $CP^{\perp}\approx{27}$~GPa, $G/B\approx{0.4}$, $\nu\approx{0.32}$), but rather low Young's moduli ($<300$\;GPa).

    \item {\bf{$\text{A}=\text{Ga}$ and $\text{M}=\{$V, Cr, W\}}} (Fig.~\ref{FIG: material selection}c). The most energetically stable compounds are $\alpha$-V$_2$GaB$_2$, V$_3$GaB$_4$, Cr$_3$GaB$_4$, $\alpha$-Cr$_2$GaB$_2$, Cr$_2$GaB$_2$, $\omega'$-W$_2$GaB$_2$, and $\gamma$-W$_2$GaB$_2$ (Tab.~\ref{Tab: best}). Although no Ga-containing MABs have been reported, several Ga-containing MAX phases (Ti$_2$GaC, Ti$_4$GaC$_3$, Cr$_2$GaC) have been explored\cite{etzkorn2009ti2gac,siebert2022synthesis}. 
    
    The here-suggested Ga-based MABs are significantly Young's-modulus- and Cauchy-pressure anisotropic. For illustration, W-containing MABs show $CP^{\perp}\geq110$~GPa and $CP^{\parallel}\leq-56$~GPa.
    
    \item {\bf{$\text{A}=\text{Ge}$ and $\text{M}=\{$V, Mn, (Re)\}}} (Fig.~\ref{FIG: material selection}d). The most energetically stable compounds are $\alpha$-V$_2$GeB$_2$ and $\alpha$-Mn$_2$GeB$_2$ (Tab.~\ref{Tab: best}). Re exhibits many competing phases. 
    Similar to Ge, also Ge-based MAB phases are currently a theoretical concept. 
    Nonetheless, there are reports on Ge-based MAX-phase thin films (Ti$_2$GeC, Ti$_3$GeC$_2$, Ti$_4$GeC$_3$, Cr$_2$GeC\cite{hogberg2005epitaxial,eklund2011epitaxial}), which may also inspire the development Ge-containing MABs.
    
    The here-suggested Ge-based MABs have generally rather low bulk moduli. The $\alpha$-Mn$_2$GeB$_2$ stand out in terms of theoretical ductility ($CP^{\parallel}\approx{84}$~GPa, $CP^{\perp}\approx{45}$~GPa, $G/B\approx{0.39}$, and $\nu\approx{0.33}$).
    
\end{itemize}

\section{Summary and conclusions}
High-throughput {\it{ab initio}} calculations were employed to facilitate rational material selection for the synthesis of novel ternary borides with MAB-phase structures.
The most promising MAB candidates were identified based on the predicted phase stability trends, energetics and volumetric proximity of competing MABs, as well as elastic-constants-based indicators of intrinsic stiffness and ductility parallel and normal to the metal/ceramic layers.

The searched chemical and phase space contained all combinations of the group 4--7 transition metals (M elements) and Al, Si, Ga, Ge or In (A elements), with 10 possible phase prototypes for each elemental combination. 
Representing the 1:1:1, 2:1:1, 2:1:2, 3:1:2, 3:1:3, and 3:1:4 M:A:B ratios, the prototypes included experimentally known MAB and MAX phases, or were inspired by typical transition metal (di)boride structures interlayered with an A layer. 
The main predictions are as follows:

\begin{enumerate}
    \item The MABs’ formation energy typically increases (indicating lower chemical stability) for M from the group 4$\to$5$\to$6$\to$7. The group of M also most notably influences the preferred phase prototype and the number of energetically- and volumetrically-close MABs, which is the highest for the group 5--6 transition metals. 
    Compared to M, the A element's effect is lower, most often decreasing stability as Al$\to$Si$\to$Ga$\to$Ge$\to$In, where In-containing MABs are also the least dense. 
    
    [{\it{Supporting data and discussion in Sec.~3.1}}] 
    
    \item Consistently with qualitative analysis of the electronic density of states, the M element significantly influences elastic properties. The strongest trend observed is a ductility increase for M from the group 4$\to$5$\to$6$\to$7. 
    Al- and Si-containing MABs typically possess the highest Young's moduli. The energetically most stable phases tend to show the lowest degree of elastic anisotropy, comparable to that of transition metal diborides.
    
    [{\it{Supporting data and discussion in Sec.~3.2 and 3.3}}]
    
    \item The suggested most promising MAB candidates combine group 5--6 transition metals and Al or Si, most often with the 2:1:2 chemistry. Based on the Cauchy pressures and Poisson's ratio, Si-based MABs are predicted to be slightly less brittle. Among them, W$_2$SiB$_2$ and Mn$_2$SiB$_2$ exhibit energetically- and volumetrically-close phases of the same chemistry, possibly facilitating transformation plasticity upon loading.
    
    [{\it{Supporting data and discussion in Sec.~3.4}}]
\end{enumerate}

In projection, our study may guide experimental development of laminated borides with optimised structure--property relationships.
Additionally, the here-produced coherent and accurate {\it{ab initio}} dataset can serve to train machine-learning models (e.g., for formation energy and elastic constants predictions) or to fit machine-learning interatomic potentials. 
Possible next steps on the computational side include (i) calculations of decomposition energies with respect to non-MAB compounds (e.g. intermetallics), (ii) the impact of point defects on phase stability and elastic properties, (iii) in-depth studies on Cr$_2$SiB$_2$ and Mn$_2$SiB$_2$ (which we suggested as promising but treated as non-magnetic), and (iv) transformation pathways between specific prototypes (e.g. for various 2:1:2 types, to assess the possibility of transformation plasticity under mechanical loads as  suggested here).

\FloatBarrier
\section*{CRediT authorship contribution statement}
{\bf{NK}}: Conceptualisation, Data curation, Formal analysis, Investigation, Methodology, Visualisation, Writing – original draft. {\bf{LH}}: Resources, Writing – review \& editing. {\bf{PHM}}: Resources, Writing – review \& editing.  {\bf{DGS}}: Conceptualisation, Methodology, Resources, Writing – review \& editing.

\section*{Declaration of Competing Interests}
The authors declare no competing interests.

\section*{Data Availability}
The data presented in this study are available from the corresponding author upon reasonable request.


\section*{Acknowledgements}
NK acknowledges the Austrian Science Fund, FWF, (T-1308). 
LH acknowledges financial support from the Swedish Government Strategic Research Area in Materials Science on Functional Materials at Linköping University SFO-Mat-LiU No. 2009 00971.
Support from Knut and Alice Wallenberg Foundation Scholar Grants KAW2016.0358 and KAW2019.0290 is also acknowledged by LH.
DGS acknowledges financial support from the Swedish Research Council (VR) through Grant Nº VR-2021-04426 and the Competence Center Functional Nanoscale Materials (FunMat-II) (Vinnova Grant No. 2022-03071).
The computations handling were enabled by resources provided by the National Academic Infrastructure for Supercomputing in Sweden (NAISS) and the Swedish National Infrastructure for Computing (SNIC) at the National Supercomputer Center (NSC) partially funded by the Swedish Research Council through grant agreements no. 2022-06725 and no.~2018-05973, as well as by the Vienna Scientific Cluster (VSC) in Austria. 

\bibliography{references}

\clearpage
\newpage
\onecolumn
\appendix
\pagenumbering{roman}
\setcounter{page}{1}
\setcounter{subsection}{0}
\renewcommand{\thesubsection}{\Roman{subsection}}
\setcounter{figure}{0}
\renewcommand{\thefigure}{S\arabic{figure}}
\setcounter{table}{0}
\renewcommand{\thetable}{S\arabic{table}}

\section*{SUPPLEMENTARY DATA}
\begin{figure}[h!t!]
    \centering
    \includegraphics[width=1\columnwidth]{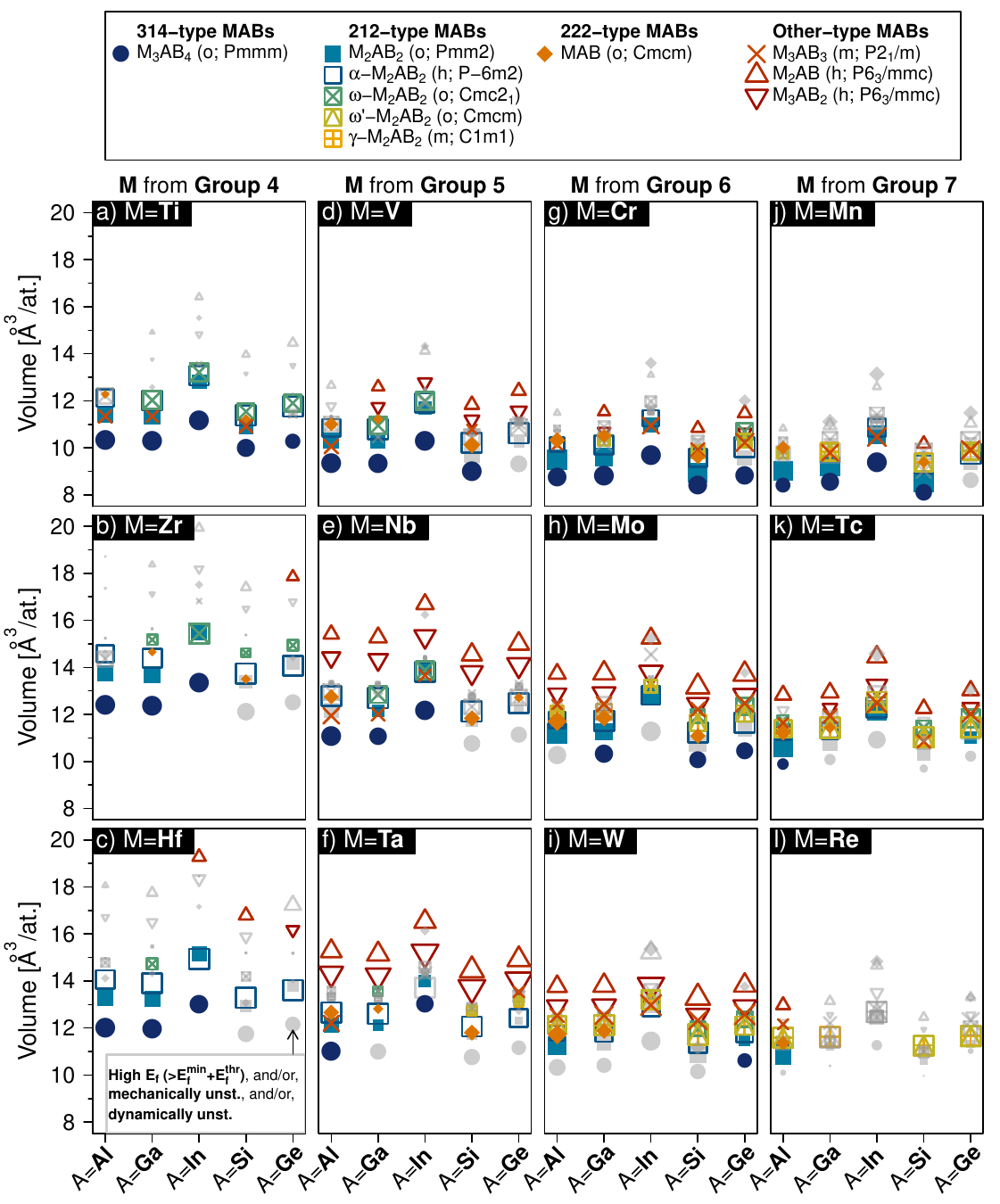}
    \caption{\small
    Trends in per-atom volume, $V_{\text{per-at.}}$, for MABs depicted in Fig.~1 in the main text. 
    MABs lying above the $E_f$ threshold ($E_f^{\text{thr}}=0.25$\;eV/at, described in the main text), and/or, mechanically, and/or dynamically unstable MABs are marked by grey colour.
    All MABs marked by colour are mechanically and dynamically stable.
    The symbol sizes scale with energetic stability quantified by the $E_f$ difference from the lowest-$E_f$ phase for a given (M, A) combination.
    }
\label{SUPPL FIG: volume}
\end{figure}

\begin{figure}[h!t!]
    \centering
    \includegraphics[width=1\columnwidth]{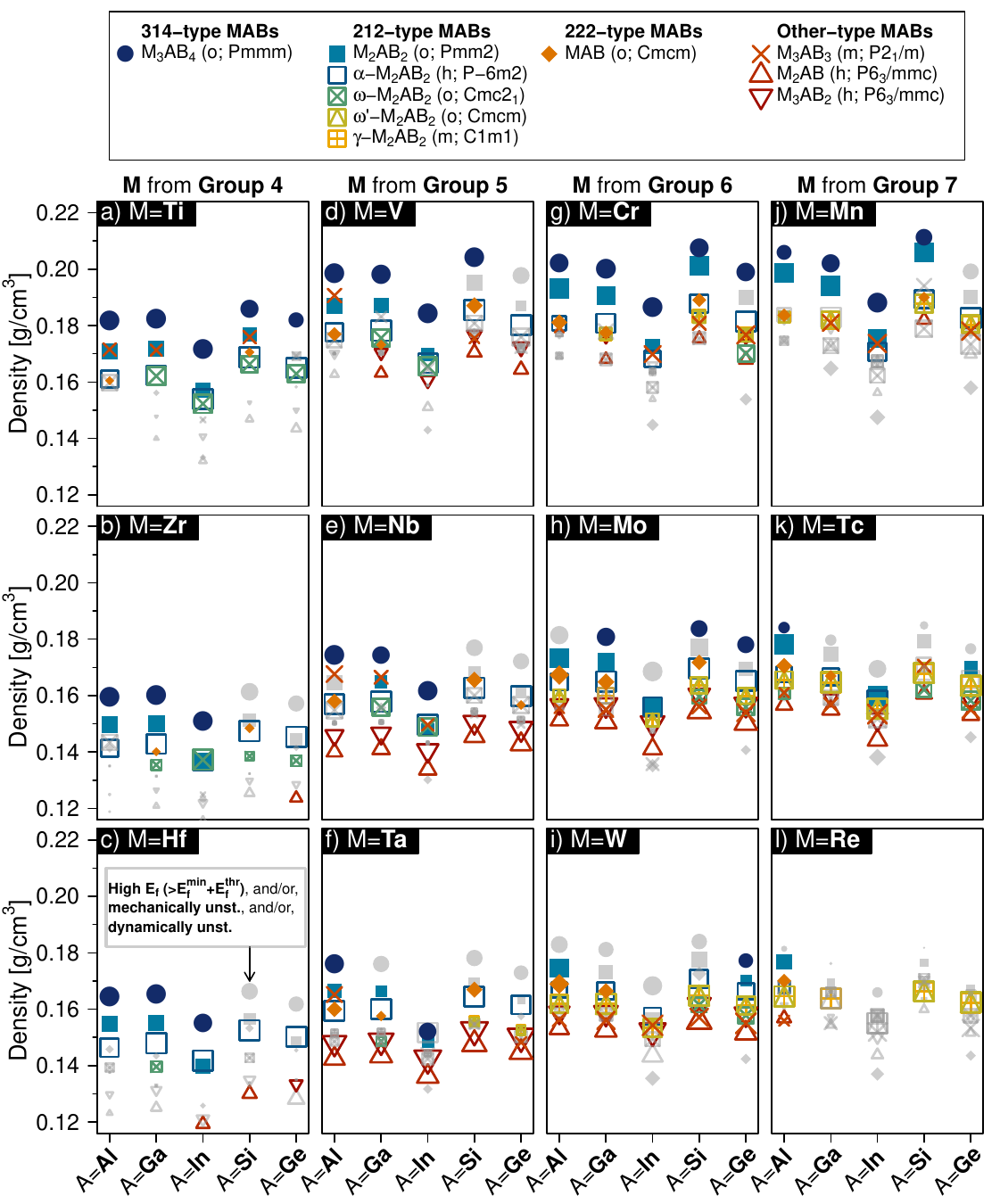}
    \caption{\small
    Trends in density, $\rho$, for MABs depicted in Fig.~1 in the main text. Denoting $V$ volume, the density is calculated as $\rho=m/V$, with $m=(m_{\text{M}}n_{\text{M}}+m_{\text{A}}n_{\text{A}} + m_{\text{B}}n_{\text{B}} )/N_A$, where $m_i$ and $n_i$ are the mass and number of atoms of element type $i$ ($i=\{$M, A, B\}), respectively, and $N_A$ is the Avogadro number. 
    MABs lying above the $E_f$ threshold ($E_f^{\text{thr}}=0.25$\;eV/at, described in the main text), and/or, mechanically, and/or dynamically unstable MABs are marked by grey colour.
    All MABs marked by colour are mechanically and dynamically stable.
    The symbol sizes scale with energetic stability quantified by the $E_f$ difference from the lowest-$E_f$ phase for a given (M, A) combination.
    }
\label{SUPPL FIG: density}
\end{figure}

\begin{figure}[h!t!]
    \centering
    \includegraphics[width=1\columnwidth]{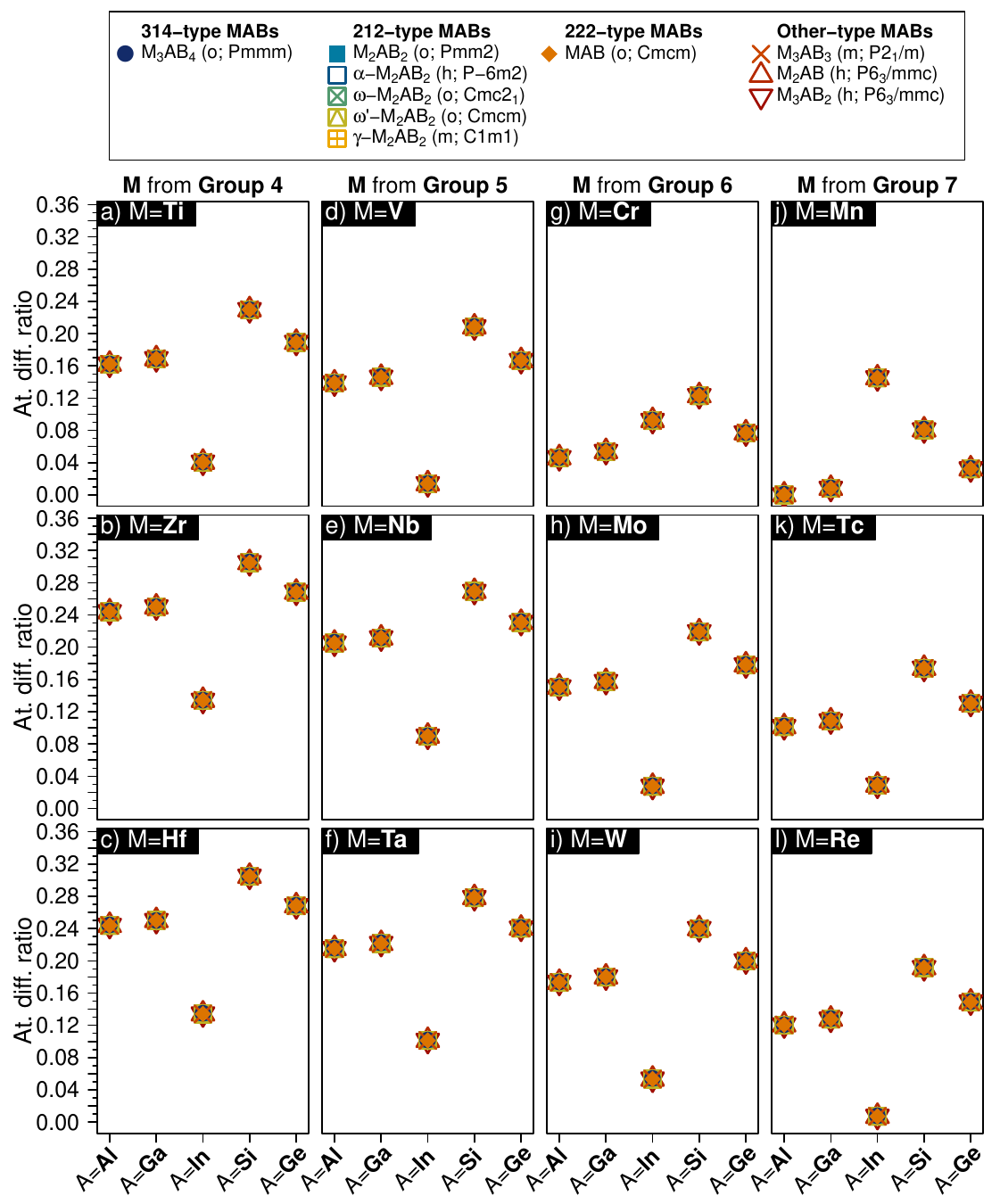}
    \caption{\small
    Trends in atomic difference ratio (alternatively called the ``size factor''), $\Delta r$, for MABs depicted in Fig.~1 in the main text. Following Zhang et al.~\cite{zhang2020role}, $\Delta r$ is calculated as $\Delta r=\frac{|r_M-r_A|}{r_M}$, where $r_M$ ($r_A$) represent atomic radius of the M (A) element in a MAB phase.
    From the definition, $\Delta r$, is the same irrespective of the phase prototype, therefore, all points for a given (M, A) combination overlap.
    }
\label{SUPPL FIG: at diff ratio}
\end{figure}

\begin{figure}[h!t!]
    \centering
    \includegraphics[width=1\columnwidth]{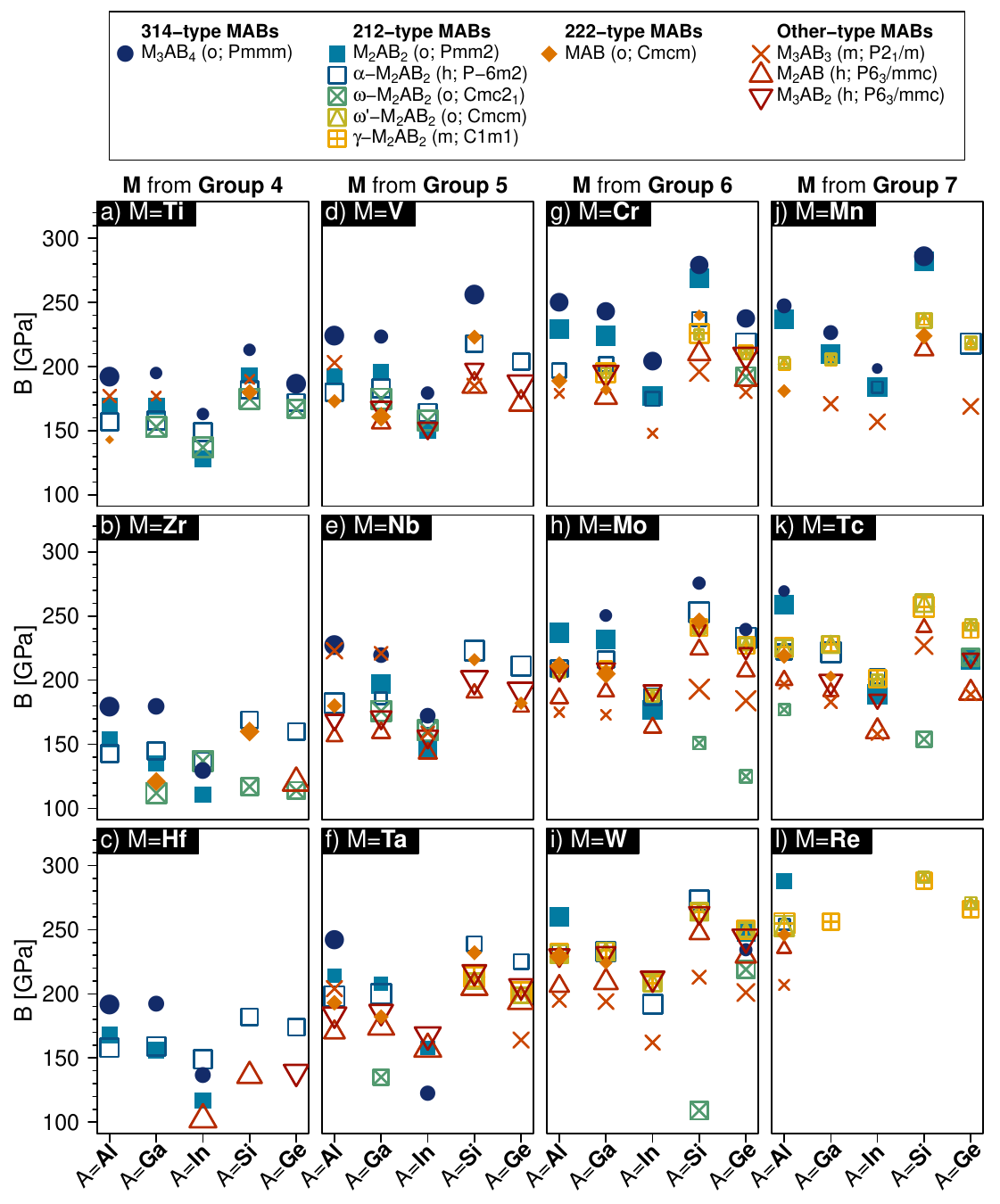}
    \caption{\small
    Polycrystalline bulk modulus, $B$, calculated for all MABs fulfilling all stability criteria ($E_f$ ``close'' above that of the lowest-energy phase, mechanical and dynamical stability).
    The symbol sizes scale with energetic stability quantified by the $E_f$ difference from the lowest-$E_f$ phase for a given (M, A) combination.
    }
\label{SUPPL FIG: B}
\end{figure}

\begin{figure}[h!t!]
    \centering
    \includegraphics[width=1\columnwidth]{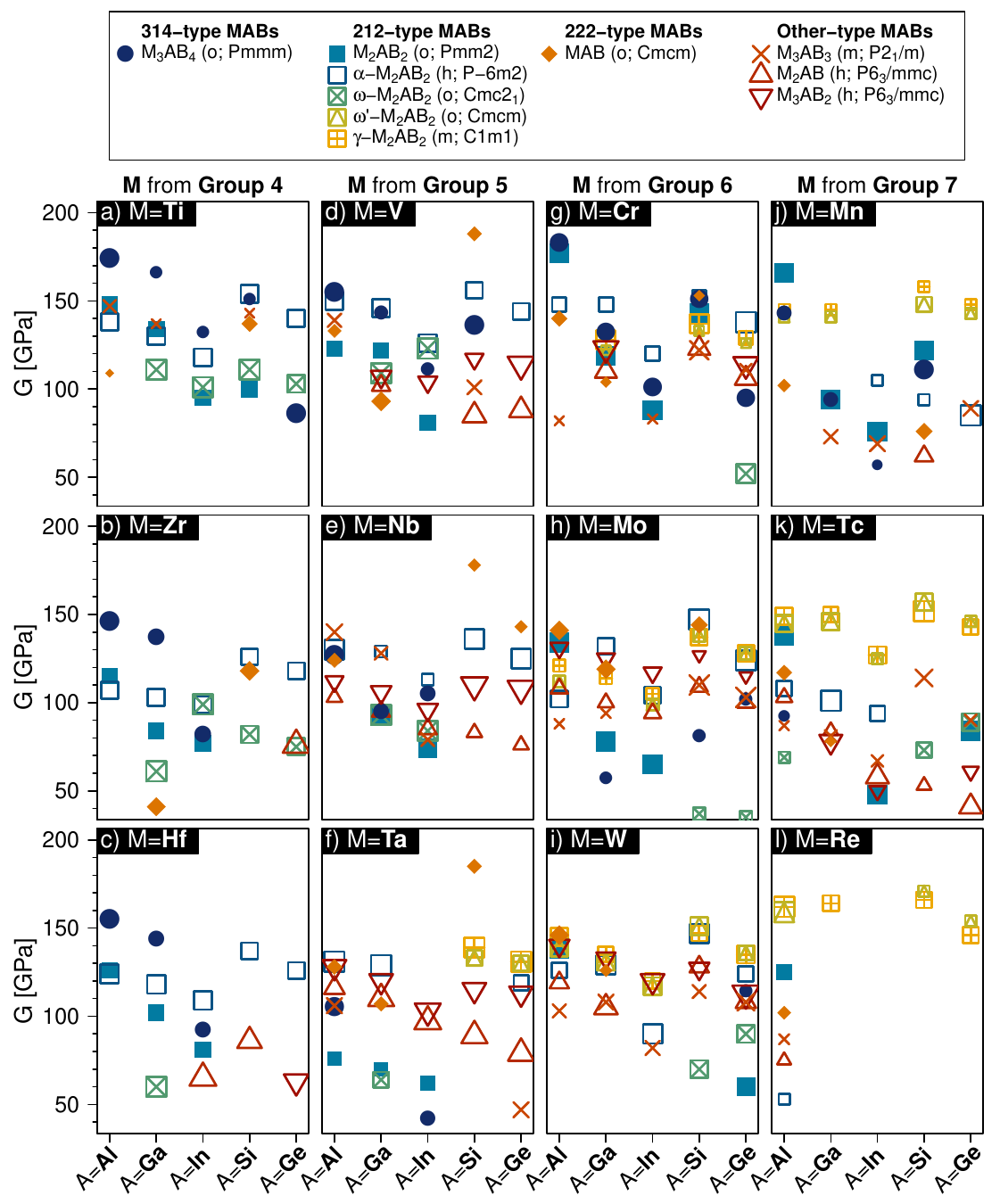}
    \caption{\small
    Polycrystalline shear modulus, $G$, calculated for all MABs fulfilling our stability criteria ($E_f$ ``close'' above that of the lowest-energy phase, mechanical and dynamical stability).
    The symbol sizes scale with energetic stability quantified by the $E_f$ difference from the lowest-$E_f$ phase for a given (M, A) combination.
    }
\label{SUPPL FIG: G}
\end{figure}

\begin{figure}[h!t!]
    \centering
    \includegraphics[width=1\columnwidth]{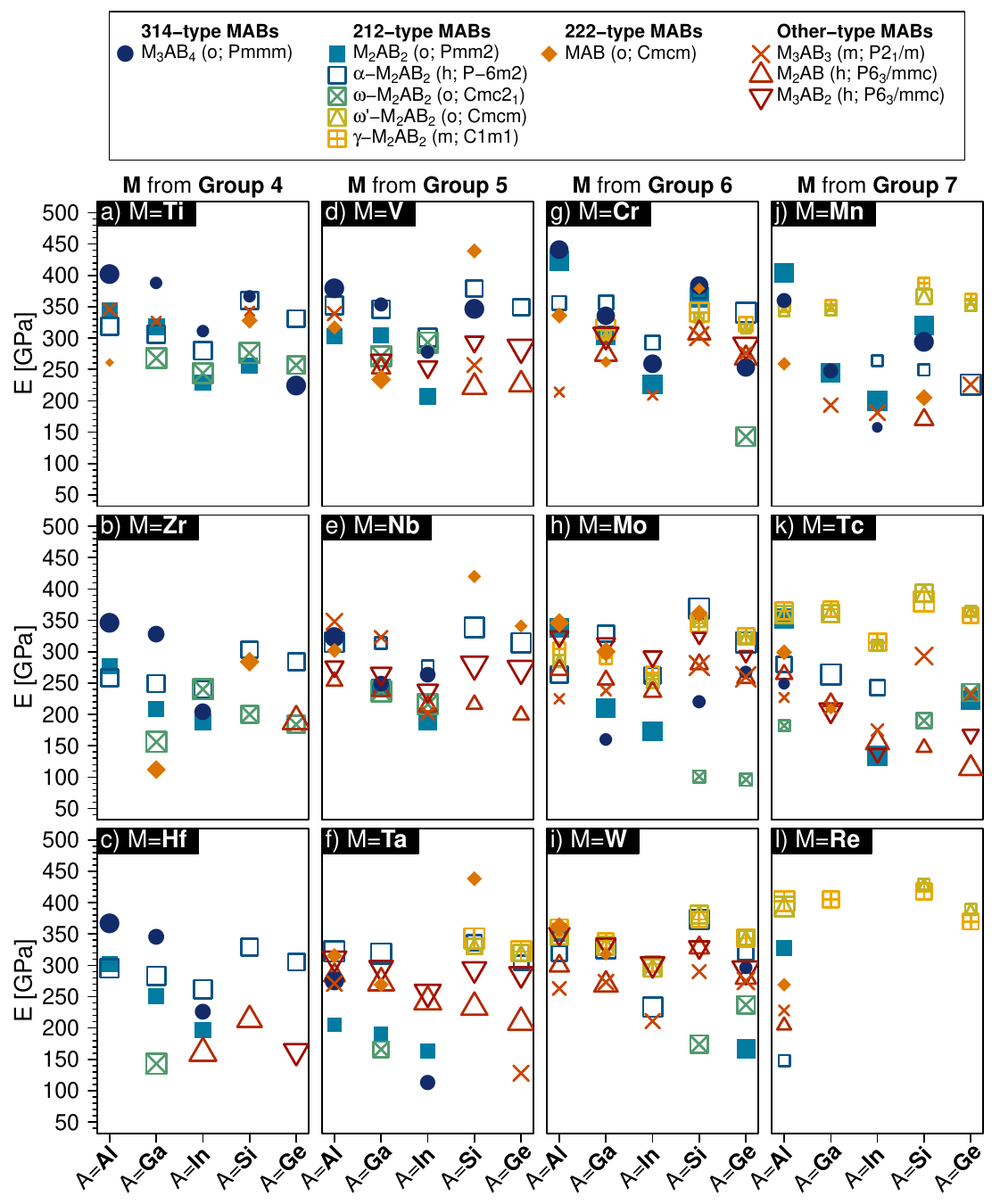}
    \caption{\small
    Polycrystalline Young's modulus, $E$, calculated for all MABs fulfilling our stability criteria ($E_f$ ``close'' above that of the lowest-energy phase, mechanical and dynamical stability).
    The symbol sizes scale with energetic stability quantified by the $E_f$ difference from the lowest-$E_f$ phase for a given (M, A) combination.
    }
\label{SUPPL FIG: E}
\end{figure}

\begin{figure}[h!t!]
    \centering
    \includegraphics[width=1\columnwidth]{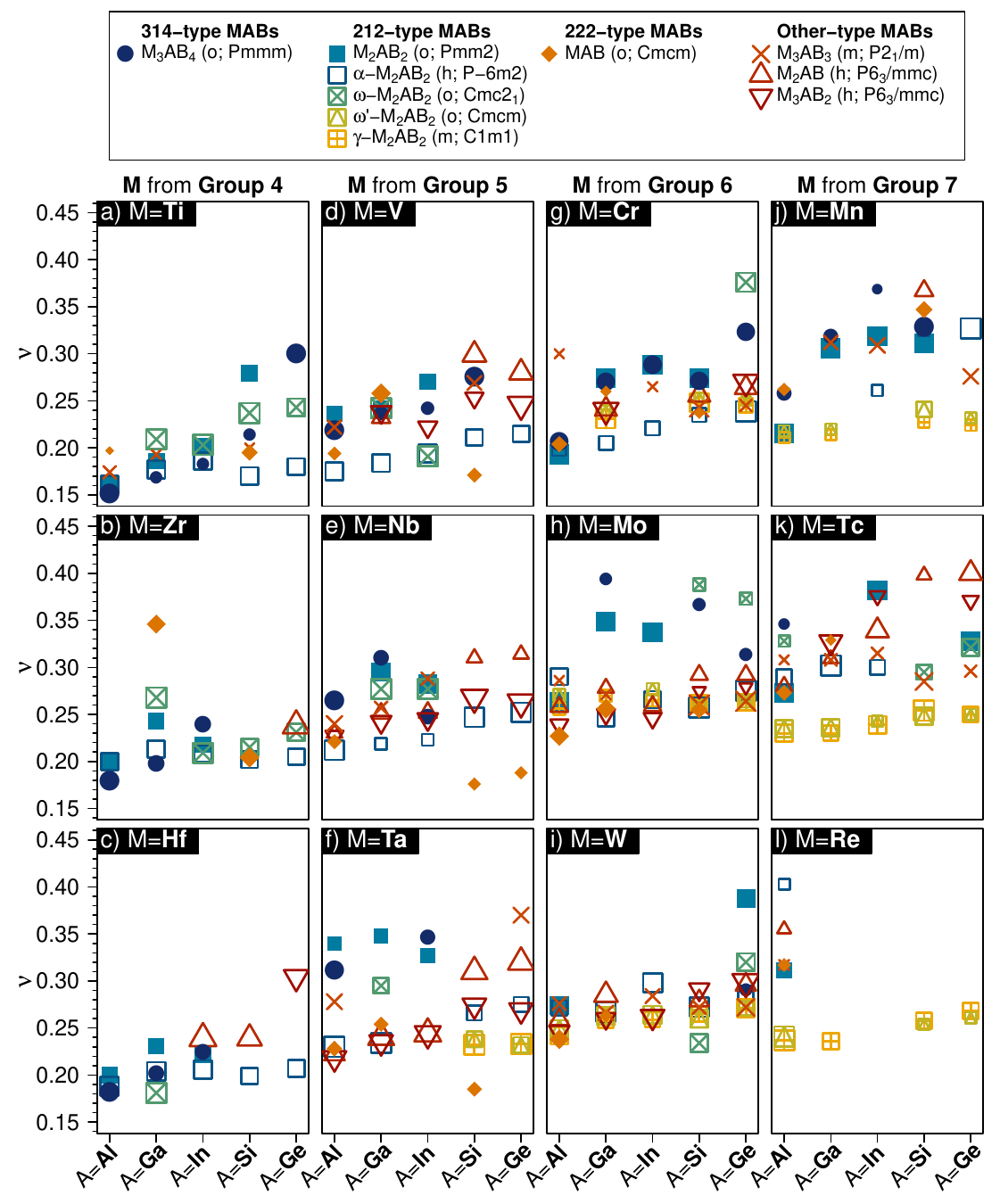}
    \caption{\small
    Polycrystalline Poisson's ratio, $\nu$, calculated for all MABs fulfilling our stability criteria ($E_f$ ``close'' above that of the lowest-energy phase, mechanical and dynamical stability).
    The symbol sizes scale with energetic stability quantified by the $E_f$ difference from the lowest-$E_f$ phase for a given (M, A) combination.
    }
\label{SUPPL FIG: nu}
\end{figure}

\begin{figure}[h!t!]
    \centering
    \includegraphics[width=1\columnwidth]{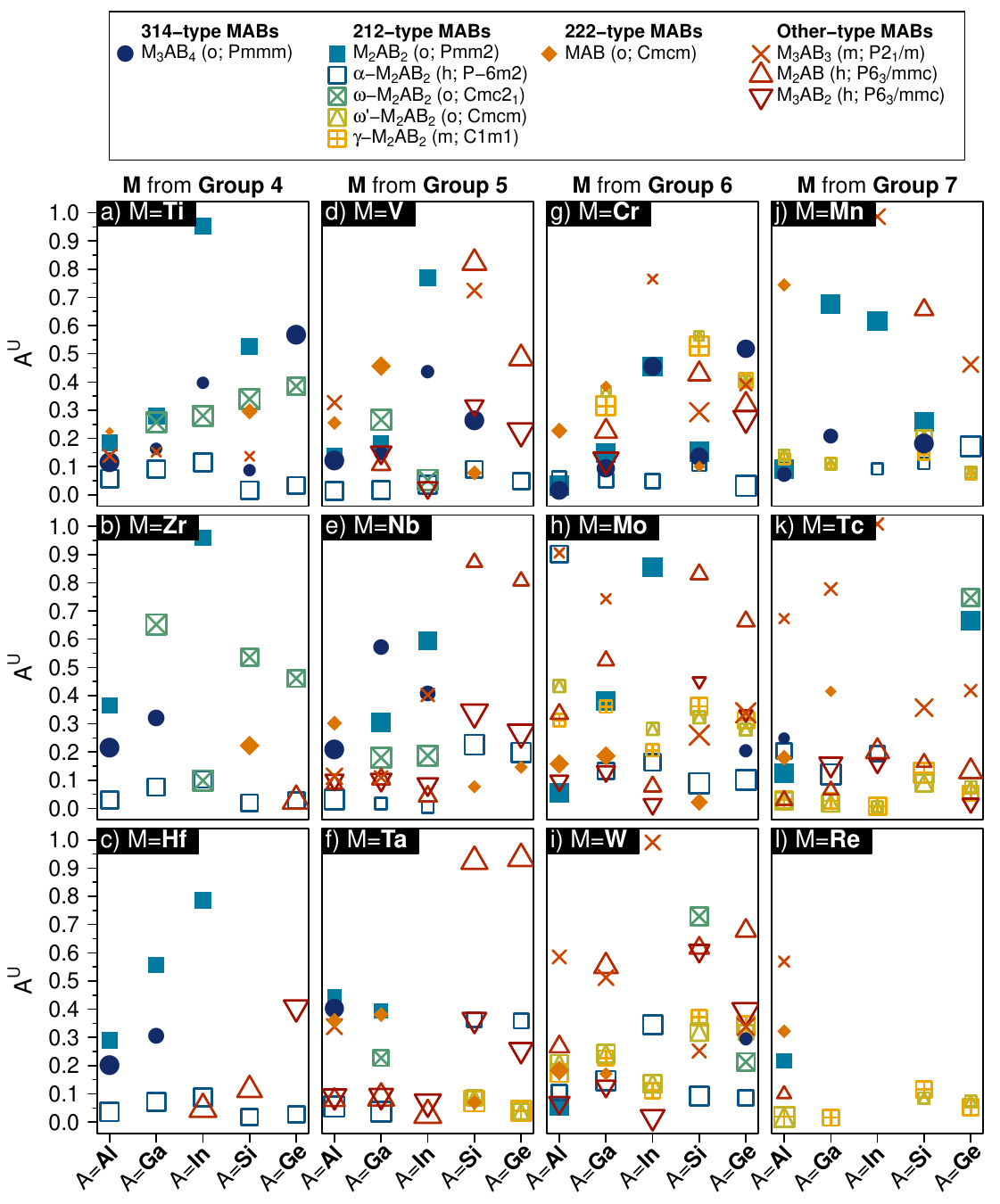}
    \caption{\small
    Universal anisotropy index, $A^{\text{U}}$ calculated for all MABs fulfilling our stability criteria ($E_f$ ``close'' above that of the lowest-energy phase, mechanical and dynamical stability). 
    The $A^U$ is calculated as $A^U=5\frac{G^V}{G^R}-\frac{B^V}{B^R}-6$ (Eq. 9 in Ref.~\cite{ranganathan2008universal}), where $G$ and $B$ are the shear and bulk moduli and the upper indexes, $V$ and $R$, denote the Voigt and Reuss estimates.
    The symbol sizes scale with energetic stability quantified by the $E_f$ difference from the lowest-$E_f$ phase for a given (M, A) combination.
    }
\label{SUPPL FIG: Anisotropy}
\end{figure}

\begin{itemize}
    \item Supplementary figures Fig.~\ref{SUPPL FIG: volume}, Fig.~\ref{SUPPL FIG: density}, and Fig.~\ref{SUPPL FIG: at diff ratio} complement the discussion of stability trends in Section 3.1 of the main text. 
    
    \item Supplementary figures Fig.~\ref{SUPPL FIG: B}, Fig.~\ref{SUPPL FIG: G}, Fig.~\ref{SUPPL FIG: E}, Fig.~\ref{SUPPL FIG: nu}, and Fig.~\ref{SUPPL FIG: Anisotropy}  complement the discussion of mechanical properties in Section 3.3 of the main text.
\end{itemize}

\end{document}